\documentclass{aa}  
\usepackage{graphicx}
\usepackage{txfonts}
\begin{document}

   \title{Characterizing the line emission from molecular clouds.
    \thanks{Table A.1. is only available in electronic form
   	at the CDS via anonymous ftp to cdsarc.u-strasbg.fr (130.79.128.5)
   	or via http://cdsweb.u-strasbg.fr/cgi-bin/qcat?J/A+A/}}
   \subtitle{Stratified random sampling of the Perseus cloud}

   \author{M. Tafalla\inst{1} \and A. Usero\inst{1} \and A. Hacar\inst{2,3}} 

   \institute{Observatorio Astron\'omico Nacional (IGN), Alfonso XII 3, 
       E-28014 Madrid, Spain \\
              \email{m.tafalla@oan.es, a.usero@oan.es}
         \and
              Leiden Observatory, Leiden University, PO Box 9513, 2300 Leiden, The Netherlands
         \and
           Department of Astrophysics, University of Vienna, T\"urkenschanzstrasse 17, 1180 Vienna, Austria \\ 
          \email{alvaro.hacar@univie.ac.at} }

   \date{Received 23 June 2020 / Accepted 8 December 2020}

% \abstract{}{}{}{}{} 
% 5 {} token are mandatory
 
  \abstract
    {The traditional approach to characterize the structure of molecular clouds 
    is to map their line emission.}
   {We aim to test and apply 
   	a stratified random sampling technique that can characterize the line emission
   from molecular clouds more efficiently than mapping.}
   {We sampled the molecular emission from the Perseus cloud using
   	the H$_2$ column density as a proxy.
   	We divided the cloud into ten logarithmically spaced column density bins,
   	and we randomly selected ten positions from each bin.
   	The resulting 100 cloud positions were observed with the IRAM 30m telescope, covering
    the 3mm-wavelength band and parts of the 2mm and 1mm bands.}
   {We focus our analysis on the 11 molecular species (plus isotopologs) 
   	detected toward most column 
   	density bins. In all cases, the line intensity is tightly correlated with the H$_2$ 
   	column density. For the CO isotopologs, the trend is relatively flat, while for
   	 high-dipole moment species such as HCN, CS, HCO$^+$, and HNC,
   	the trend is approximately linear. 
   	To reproduce this behavior, we developed a cloud model in which the gas density
   	increases with column density, and where most species
   	have abundance profiles characterized by an outer photodissociation edge and an 
   	inner freeze-out drop. 
    With this model, we determine that the intensity behavior of the high-dipole moment species
    arises from a combination of excitation effects and molecular freeze out,
    with some modulation from optical depth.
    This quasi-linear dependence with the H$_2$ column density 
    makes the gas at low column densities dominate the cloud-integrated emission.
    It also makes the emission from most high-dipole moment species proportional to the cloud mass inside the photodissociation edge.}
   {Stratified random sampling is an efficient technique for characterizing 
   	the emission 
   	from whole molecular clouds. When applied to Perseus, it shows that 
   	despite the complex appearance of the cloud, the molecular emission follows a relatively 
   	simple pattern. A comparison with available studies of whole clouds
   	suggests that this emission pattern may be common.
   }


   \maketitle
%
%-------------------------------------------------------------------

\section{Introduction}

Molecular clouds are the coldest and densest constituents of the interstellar
medium and harbor in their interiors the sites where stars are born.
Clouds are believed to
form and evolve by the complex interplay between
turbulence, gravity, and magnetic fields, although the exact role
of each factor is still a matter of debate
(see \citealt{hen12,dob14} for recent reviews).
Progress in our understanding of clouds requires characterizing their
physical and chemical structure and interpreting this structure in terms of
the different forces acting on the gas.
This characterization is usually done
by mapping the emission from molecular lines since these lines
provide unique information on the density, temperature, kinematics,
and molecular composition of the cloud gas \citep{eva99}.

The large angular size of the nearby clouds makes
mapping their line emission 
very time consuming, especially for the weak subthermally excited 
lines that are most sensitive to the physical conditions of the gas.
Due to this limitation, large-scale maps of clouds are almost exclusively made in
the bright and thermalized (often optically thick) lines of the CO isotopologs
(e.g., \citealt{gol08,buc10,ume17}), while the mapping 
of the more informative subthermal 
lines is restricted to the densest 
parts of the clouds (e.g., \citealt{san12,jac13}).
As a result, our view of the line emission from molecular clouds
is spatially limited
and often restricted to a small number of molecular species.

Over the past several years, a large effort has been made 
to overcome previous observing limitations  and 
map entire clouds using multiple line tracers.
This effort has been made possible by a new generation of
heterodyne receivers with large frequency bandwidths (e.g., \citealt{car12})
and has provided a first 
multiline
view of full or sizable parts of several molecular clouds.
An example of this effort is the IRAM Large Program ORION-B \citep{pet17},
which has mapped most of the Orion~B cloud in the 3mm-wavelength band,
and whose results are currently  being published \citep{ork17,gra17,bro18, ork19}.
Using a similar approach, 
\cite{wat17} carried out multiline large-scale mapping of
the high-mass star-forming region W51 with
the Mopra telescope.
Complementing this effort, \cite{kau17} compiled multiple observations of 
the Orion~A cloud made with the FCRAO telescope and used the resulting
dataset to study how the emission from the different molecular 
species originates in
different parts of the cloud. 
Although these efforts are encouraging, 
the large investment of observing time required to map      
individual clouds in multiple lines 
(often hundreds of hours, see \citealt{ork19})
suggests that full-cloud mapping will 
remain for some time a niche approach limited to the study of selected targets.

While fully mapping clouds is necessary to characterize their emission
in an unbiased way, these same observations show that clouds tend to
present a common underlying behavior despite their diverse and chaotic appearance. 
Examples of this behavior are the
Larson's relations between global
cloud properties such as mass, size, and velocity dispersion \citep{lar81,hey04},
the almost universal fractal dimension found
using area-perimeter measurements \citep{baz88,fal91},
and the common behavior of the probability distribution function of
column densities \citep{kai09,lom15}.
These and other trends suggest that most
clouds have, to first order, a common physical 
and chemical structure that could be described using a small number of parameters.
If this is so, the emission from the clouds
will also likely present a systematic behavior, of course modulated by the
characteristics of each individual system.

If clouds emit according to some simple and general pattern, it should be possible  
to characterize this pattern using a limited set of observations
instead of having to map the emission in detail. In this paper, we explore
this possibility by using a relatively sparse 
sampling technique, which we applied to the nearby \object{Perseus cloud}.
This cloud was chosen for having a number of favorable characteristics
and a significant amount of previous data (see \citealt{bal08} for a  
detailed review of the cloud properties).
It is nearby, with a distance that has been variously estimated
as 234~pc from VLBI VERA observations \citep{ver20} and 
about 300~pc from VLBA and Gaia measurements \citep{ort18,zuc19},
and has an estimated mass of $2\times 10^4~M_\odot$ (\citealt{zar16}, 
assuming a distance of 240~pc).
It is an active star-forming region that contains the young cluster 
IC~348 \citep{str74,lad06}, the embedded cluster NGC~1333 \citep{str76,gut08},
and a more distributed population of young stars and protostars \citep{jor07,reb07}.

The large-scale emission of the different CO isotopologs in Perseus has been mapped by
\cite{bac86}, \cite{ung87}, \cite{rid06}, \cite{sun06}, and \cite{cur10}, 
while the dust component has been mapped using mm, submm, and FIR
emission \citep{hat05, eno06, che16,pez20} and optical and IR extinction 
\citep{bac86,sch08,lom10,zar16}.
In addition to these large-scale studies, a number 
of observations targeting 
the dense cores have been presented by \cite{lad94},
\cite{kir06}, \cite{ros08a}, and \cite{hac17} among others.
All these studies (and others not mentioned here for brevity)
make Perseus one of the best studied clouds, and therefore an ideal 
region to test a sampling technique. 

In this paper,  we present an
emission survey of Perseus with two distinct goals:
(1) to test whether the cloud emission can be characterized using a sampling technique and,
assuming that the answer is positive, 
(2) to use the sampling data to reconstruct the cloud emission and
infer the gas physical and chemical properties.
Since the focus of our study is the intensity of the line emission, our sampling
technique (described in the next section) is designed to highlight this parameter 
at the expense of others. As a result, our analysis will not touch upon other important 
cloud properties, such as the gas velocity field, which require a different approach
for their study.

\section{Stratified random sampling}
\label{sec-_sampling}

\begin{figure*}
    \centering
    \includegraphics[width=\hsize]{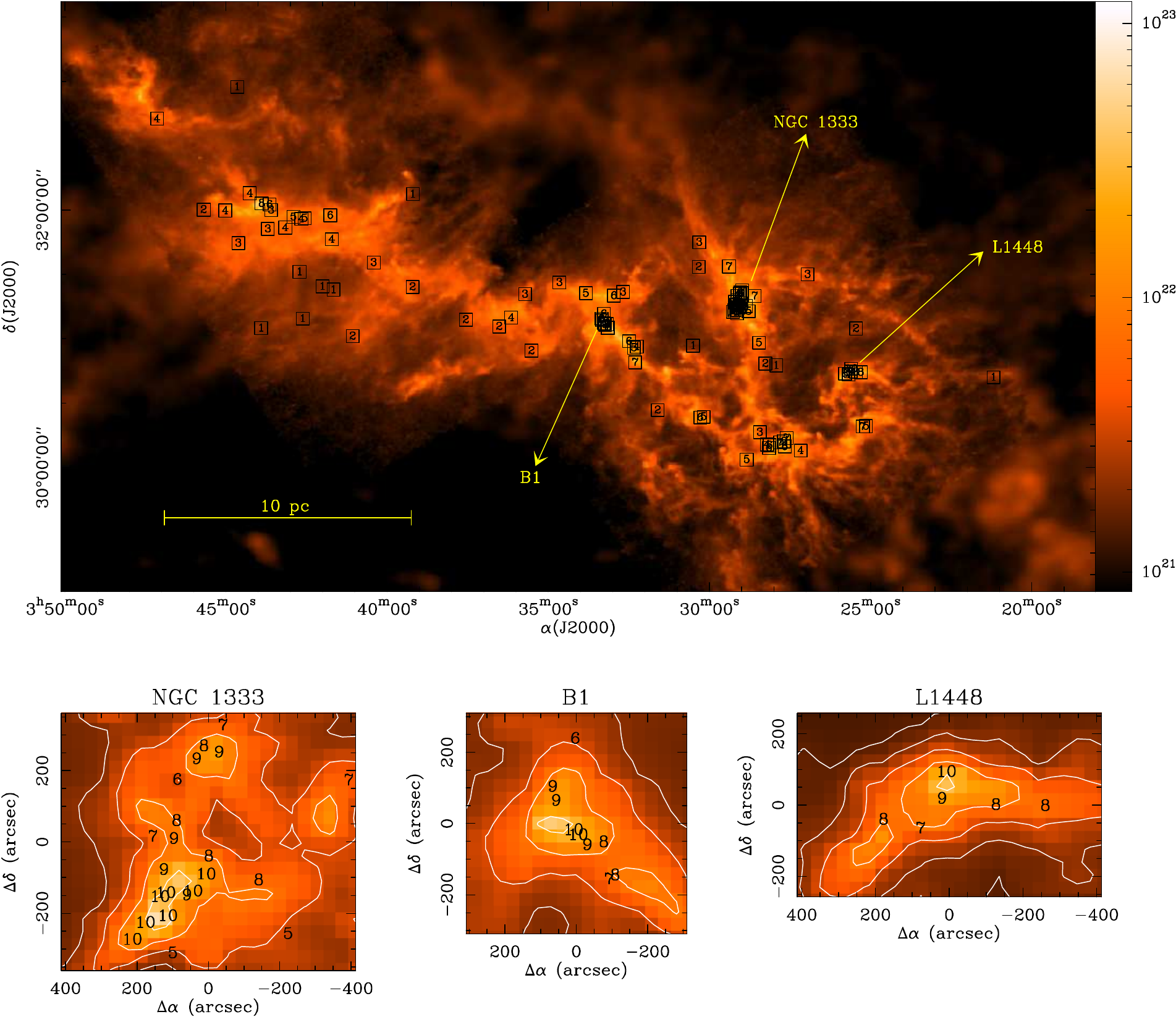}
    \caption{Sampling of the Perseus cloud.
    	{\em Top:} H$_2$ column density map of the Perseus molecular cloud from \cite{zar16}
    	with squares indicating the location of the 100 positions selected to characterize the line emission from the cloud
    	using stratified random sampling.
    	The number in each square indicates the column density bin to which the position belongs,
    	and ranges from 1 (lowest column density) to 10 (highest).
    	{\em Bottom:} Expanded view of the three regions that concentrate most of the high column density
    	positions, and that correspond to the well-known star-forming regions NGC 1333, B1, 
    	and L1448.
    	Contours are in units of $N$(H$_2$), start at $1.25 \times 10^{21}$~cm$^{-2}$, and increase by factors	of 2.5. 
    	The 10~pc scale bar assumes a distance of 300~pc.}
    \label{map}
\end{figure*}

The goal of our work is to sample
the different regimes of the cloud molecular emission 
using a relatively small number of positions,
even if the cloud emission properties are not initially well known.
Choosing positions at random does not seem like a good strategy,
since a typical cloud has so many more positions with weak emission
that the probability of randomly picking bright positions is negligibly low
\citep{ros08b}.

A better option is to guide the choice of positions by a proxy
that is expected to correlate with the molecular emission.
A natural choice for this proxy is the H$_2$ column density,
which has been shown by principal component analysis to
have a dominant contribution to the emission of some clouds \citep{ung97,gra17}.
The H$_2$ column density, in addition,
can be determined with accuracy over entire
molecular clouds using a combination of dust extinction and
emission measurements (e.g., \citealt{lom14}).
For the Perseus cloud, \cite{zar16} have recently determined
the distribution of H$_2$ column density combining
dust emission data from the {\em Herschel} and {\em Planck}
satellites together with NIR dust extinction measurements from the 2MASS survey.
This determination covers the full extent of the cloud and has a dynamic range 
of about two
orders of magnitude. In addition, it has 
an angular resolution of $36''$, which is similar to what is currently
achievable with single-dish radio telescopes.

To sample the cloud using the H$_2$ column density as a proxy,
we need to 
sample the distribution of H$_2$ column densities in the cloud so 
that each column density regime is well represented.
Properly sampling this distribution, however,
requires some care since, like the emission, the population
of column densities
is overwhelmingly dominated by the positions
with the lowest values. \cite{zar16} found that
the probability distribution function of the column density follows a 
power law with a slope of $-3$, and as a result, that the number of positions
at the low end of the distribution
($N(\mathrm{H}_2) \approx 10^{21}$~cm$^{-2}$) exceeds the number of positions at the high end
($N(\mathrm{H}_2)\approx 10^{23}$~cm$^{-2}$) by about six orders of magnitude (Fig.~8
in \citealt{zar16}). Sampling the distribution by choosing positions
with random column density is therefore impractical
since it would require selecting about one million samples to ensure 
that the full range of column densities is covered.

\begin{figure}
    \centering
    \includegraphics[width=\hsize]{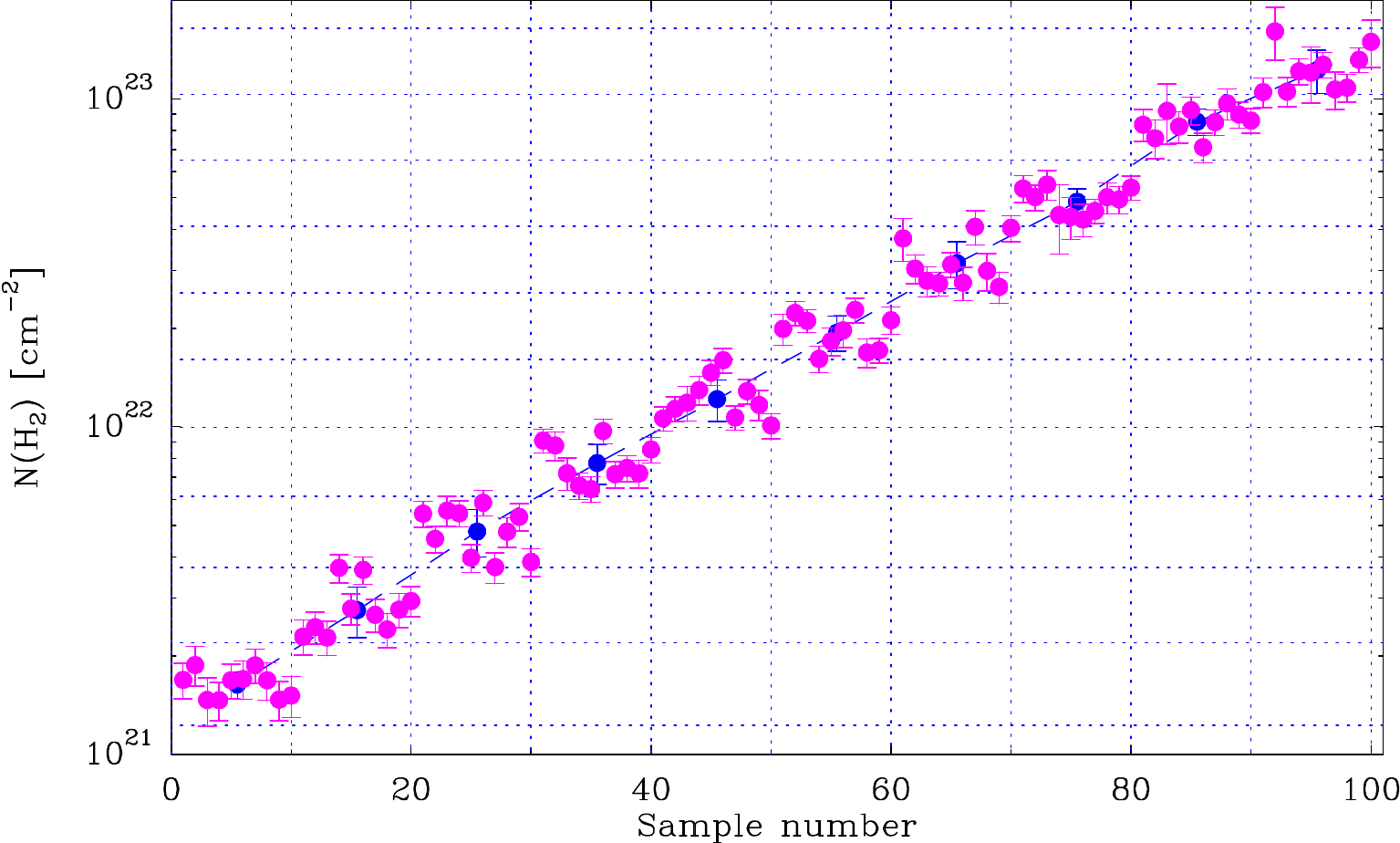}
    \caption{H$_2$ column density as a function of sample number for the 100 positions randomly
    	chosen to represent the Perseus molecular cloud (magenta circles). The dashed horizontal 
    	lines mark the boundaries of the ten column-density bins in which the cloud was divided, and
    	the vertical lines enclose the ten points chosen to sample each bin. The blue circles
    	indicate the geometrical mean of the H$_2$ column densities inside each bin, and their
    	error bars indicate their dispersion. All values derived from \cite{zar16} data.}
    \label{sampl}
\end{figure}

A more practical approach is
to first identify the different column density regimes in the cloud and 
then to sample each regime by choosing a number of cloud positions at random.
This sampling approach is an instance 
of the ``stratified random sampling''
technique often used in surveys \citep{coc77}, and whose name derives
from the word ``strata'' used to denote the
different subpopulations of the sample, which in our case correspond to the 
column density regimes of the cloud.

Since the distribution of column densities in Perseus follows a power law 
over the approximately two orders of magnitude for which the extinction
measurements are reliable ($1.5 \times 10^{21}$ < $N$(H$_2$) < $1.5 \times 10^{23}$~cm$^{-2}$, 
corresponding to 0.2 < $A_{\mathrm K}$ < 20~mag, \citealt{zar16}), 
we chose to bin this range 
using logarithmically spaced column-density intervals.
The width of these intervals was taken as
0.2 dex, equivalent to a factor of 1.6 
in column density, with the expectation that the molecular emission will not change
dramatically (more than a factor of 2)
between the bins. As shown below, this expectation is satisfied over most of the cloud,
although it seems to break down toward the lowest column density bin
($A_{\mathrm V} \approx 1-2$~mag), where the emission
often drops precipitously due to photodissociation at the cloud edge.
Our column density sampling is clearly not fine enough to resolve the details of the 
cloud boundary, but seems to work well over the rest of the cloud.
Since the cloud column density range spans two orders 
of magnitude, a choice of a 0.2~dex
interval resulted in a total of ten logarithmically spaced column-density bins.

In standard applications of the stratified sampling method, 
the strata are sampled proportionally to their population.
As mentioned above, this approach
is impractical when sampling the
column density bins of Perseus (or any other cloud) because their 
populations can differ by multiple orders of magnitude.
As an expedient solution motivated by the limited observing time available,
we chose to sample each bin using the same number of positions, which we
took as ten. This number was chosen as 
a compromise between the competing needs of
observing enough positions per bin
to determine with some accuracy the mean and the dispersion of the intensity,
and the need to make long enough integrations
to detect the weak emission of the positions in the lowest column density bins.
Since the success of this sampling strategy depends strongly on the intensity and 
the dispersion of the emission inside the bins, which was not known before the observations,
our choice of sampling parameters should be taken only as a tentative initial approach.
A proper optimization of the stratified random sampling technique to characterize
molecular clouds is still needed to assess  a number
of possible weaknesses of our approach, which include the use of limited
sampling in the very extended low column density bins (which could lead
to statistical fluctuations)
and the use of H$_2$ column density as the sole guide to sample the cloud emission
to the exclusion of other parameters, such as the gas temperature or the
star-formation activity.
Carrying out this investigation requires having full maps of clouds
in a variety of molecular tracers, which fortunately  is now becoming possible 
thanks to the efforts of dedicated observing programs such as ORION-B \citep{pet17} and 
LEGO \citep{bar20}.

Once the number of sampling positions had been determined, the
practical choice of selecting them at random
was done by first identifying all the pixels in the extinction map of \cite{zar16} that belong to 
each column density bin. These pixels were listed 
in a table, and ten of them were selected by repeatedly drawing 
random numbers uniformly distributed between one and the number of pixels
in the bin. Table~\ref{tbl_master} presents a list with the coordinates of the resulting 
100 cloud positions. Their relative location inside the cloud is shown 
in Fig.~\ref{map}.

To illustrate the two orders of magnitude in H$_2$ column density covered by the sample,
Fig.~2 shows the value determined using the prescription from \citealt{zar16}
for each sample position (magenta symbols). Each column density
bin is enclosed between dashed horizontal lines, and as expected, 
the points inside each bin are distributed randomly
in column density.
The figure also shows the geometrical mean and dispersion inside each bin
(blue symbols). We note that the error bars in the individual column densities are typically
smaller than the dispersion inside the bins, although an
additional source of uncertainty in the column density arises from the particular
choice of extinction per H atom assumed by \cite{zar16}, which could shift the estimates
globally by a factor of up to about 1.5 depending on the true dust properties 
(Fig.~2 in \citealt{dra03}).

\section{Observations}
\label{sect_obs}

We observed our sample of Perseus positions using the 
Institut de Radioastronomie Millim\'etrique (IRAM)
30m-diameter telescope in Pico Veleta (Spain) during three runs 
in June 2017, September 2017, and January 2018.
In the first two runs, the 3mm channel of the Eight MIxer Receiver (EMIR, 
\citealt{car12}) was used to
observe the full frequency range from 83.7 and 115.8~GHz
(telescope FWHM of $29''$ to $21''$).
These observations used the facility fast Fourier Transform Spectrometer
(FTS, \citealt{kle12}), which was configured
to cover the receiver instantaneous passband with a 
frequency resolution of 200~kHz ($\approx 0.6$~km s$^{-1}$).

For all positions,
the integration time was approximately 10 minutes after combining 
the two linear polarizations, except for 
the positions in the lower column density 
bin, which were observed twice as long to compensate for their weaker lines.
Additional observations of 33 positions from our sample were carried out 
using simultaneously the 3mm and 2mm channels of EMIR and covering
two narrow frequency ranges centered on HCO$^+$(1--0) and CS(3--2).
These observations had the goal of obtaining high velocity resolution
spectra, and used as a backend the Versatile SPectrometer Array 
(VESPA) with
a frequency resolution of 20~kHz ($\approx 0.06$~km s$^{-1}$).

In the last observing run (January 2018), we used 
the 1mm channel of the EMIR receiver 
to observe four frequency windows within the range 213.7-267.7~GHz 
(telescope FWHM of $11''$-$9''$). These windows were
selected for containing higher-$J$ transitions of some 3mm target lines, such as 
CO(2--1), HCN(3--2), and CS(5--4). 
Again, the FTS backend was used
to cover as much passband as possible with a frequency resolution of 
200~kHz, equivalent to $\approx 0.25$~km s$^{-1}$ at the frequency of operation.
Since the 1mm lines are weaker than those at 3mm, the
lowest three column density bins of the sample were not fully observed, and
the integration time per point typically
ranged between 10 and 40 minutes (after combining polarizations),
depending on the strength of the line.

All observations were carried out in frequency switching 
mode with symmetric offsets of $\pm 7.7$~MHz. Atmospheric calibration 
was performed every 10-15 minutes using the standard sky-ambient-cold 
load cycle, and the telescope pointing and focus were checked and 
corrected every two hours approximately. The resulting folded spectra  
were further processed using the CLASS 
program\footnote{\url{http://www.iram.fr/IRAMFR/GILDAS}},
with which
repeated observations and overlapping frequency windows were {\tt STITCH}ed 
together. Polynomial baselines were used to remove ripples in the passband,
and the intensity was  converted to the main beam brightness scale using the
facility-provided telescope efficiencies. 
Typical rms in the spectra range from 6 to 9 mK per 0.6~km~s$^{-1}$ channel at 100 GHz.

\begin{figure*}
	\centering
	\includegraphics[width=\hsize]{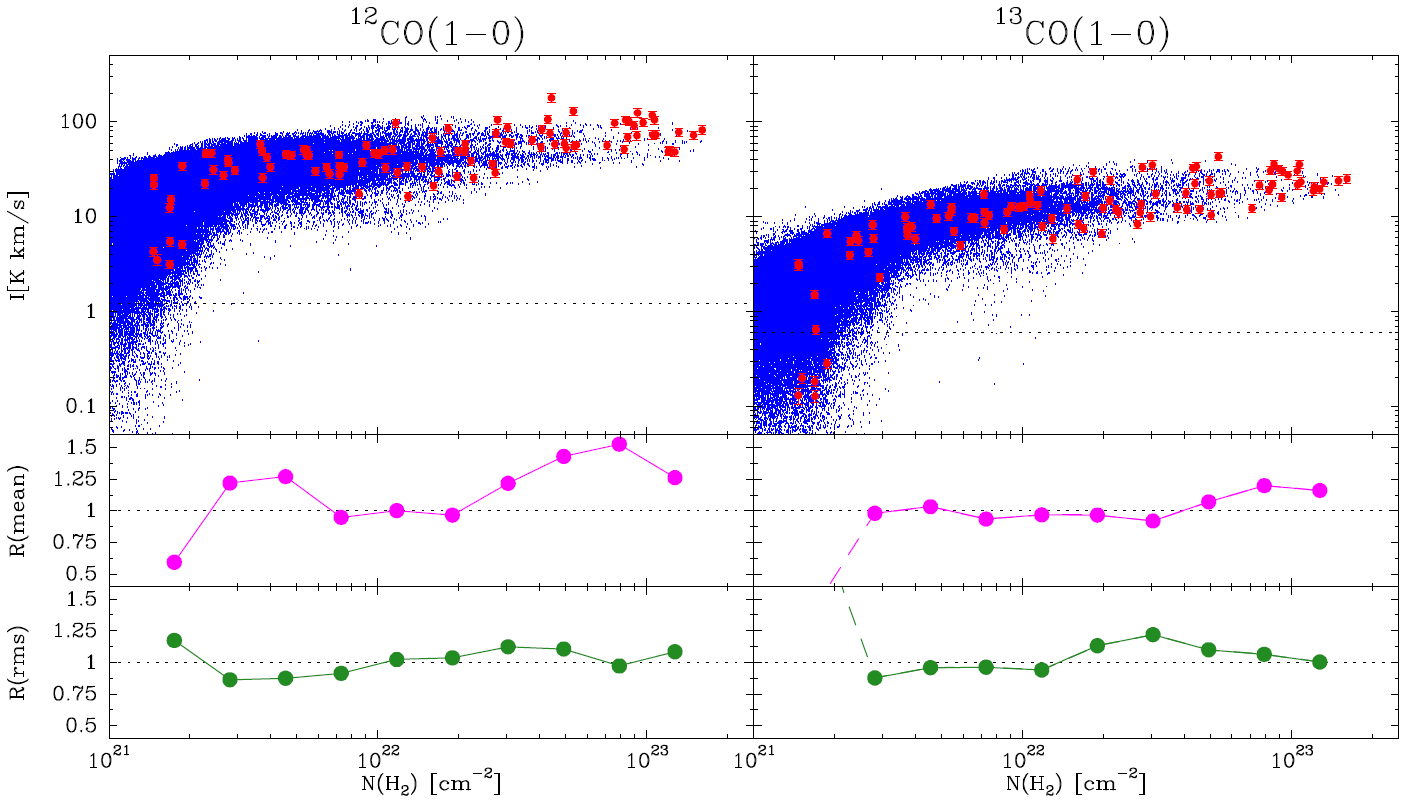}
	\caption{Comparison between sampling observations and COMPLETE full maps for $^{12}$CO(1--0) and $^{13}$CO(1--0).
		{\em Top panels: } Scatter plots of the integrated intensity  
		as a function of $N$(H$_2$).
		The blue dots represent the results from the COMPLETE maps presented by \cite{rid06} and
		each of them contains more than $2 \times 10^5$ spectra. The red symbols represent the
		100 measurements obtained using stratified random sampling.
		The dotted lines mark the $3 \sigma$ detection level of the COMPLETE maps.
		{\em Middle panels: } Ratio between the mean intensity derived using the sampling 
		data and the COMPLETE maps for each of the ten $N$(H$_2$) bins in which the cloud 
		was divided. For $^{13}$CO(1--0), the measurement in lowest $N$(H$_2$) bin
		is not reliable
		since the sampling data show that the intensity often lies below the COMPLETE 
		detection limit.
		{\em Bottom panels: } Ratio between the rms derived using the sampling 
		data and the COMPLETE maps. As with the mean values, the $^{13}$CO(1--0) 
		measurement for the lowest $N$(H$_2$) bin lies below the detection limit, so  
		the rms estimate is not reliable. 
		We note how both
		the mean and rms ratios for the two CO isotopologs 
		lie in the vicinity of 1, indicating that the sampling method 
		allows estimating the main emission parameters with an accuracy of
		about 50-20\% depending on the parameter.}
	\label{fig_vs_complete}
\end{figure*}

In most of the following analysis, we rely on the
integrated intensity of the different molecular lines.
These intensities were estimated from the reduced spectra
by integrating the emission over the velocity channels where a visual
inspection showed signal. For spectra with no clear signal, the
intensity was estimated by integrating the emission inside the velocity
range where the $^{13}$CO(1--0) line was detected since this abundant species was
identified toward almost all cloud positions,
and its velocity range was found to
coincide with that of all the other lines in case of mutual detection.
The uncertainty of each integrated intensity 
was estimated from the rms level of the spectrum. 
Following previous IRAM 30m studies, we
added in quadrature a 10\% calibration error to include uncertainties in 
the beam efficiencies and day-to-day variations \citep{pet17,jim19}.
Table~\ref{tbl_master} summarizes the derived intensities.

\section{Survey results}

\subsection{Comparison with the COMPLETE project}

Before analyzing the data, we test how well our sampling observations recover the main 
properties of the cloud emission.
For this, we compare our $^{12}$CO(1--0) and $^{13}$CO(1--0) intensities 
with the results from the
Coordinated Molecular Probe Line Extinction and Thermal Emission (COMPLETE) project,
which mapped the entire Perseus cloud in $^{12}$CO(1--0) and $^{13}$CO(1--0) using the FCRAO 14m telescope. 
These data
have been presented by \cite{rid06}, and their correlation with the extinction determined
using 2MASS data has been studied in detail by \cite{pin08} and \cite{goo09}.
To compare our survey data with the COMPLETE results, we first converted the COMPLETE intensities into the main beam
brightness scale using the efficiencies recommended by \cite{rid06}. We then 
resampled
the extinction map from \cite{zar16} to the same spatial grid used by COMPLETE, in order to obtain
for each position of the COMPLETE map 
an estimate of the H$_2$ column density $N$(H$_2$). This estimate 
was determined from the extinction data 
using the conversion factors recommended by \cite{zar16}.

In the top panels of Fig.~\ref{fig_vs_complete}, we represent the intensity of the 
$^{12}$CO(1--0) and $^{13}$CO(1--0) lines
as a function of the H$_2$ column density for both the sampling and COMPLETE datasets.
This type of plots constitutes the basis of most of our Perseus analysis  
presented below, so it represents the most adequate tool to perform the 
data comparison.
The COMPLETE data are represented with blue symbols and consist of 
more than $2 \times 10^5$ points, one for each spectrum used to generate
each COMPLETE map \citep{rid06}.
Superposed with red circles, we present the results from our sampling 
observations, also in main-beam brightness temperature scale and 
with the H$_2$ column density estimated
using the \cite{zar16} prescription. 

\begin{figure*}
	\centering
	\includegraphics[width=\hsize]{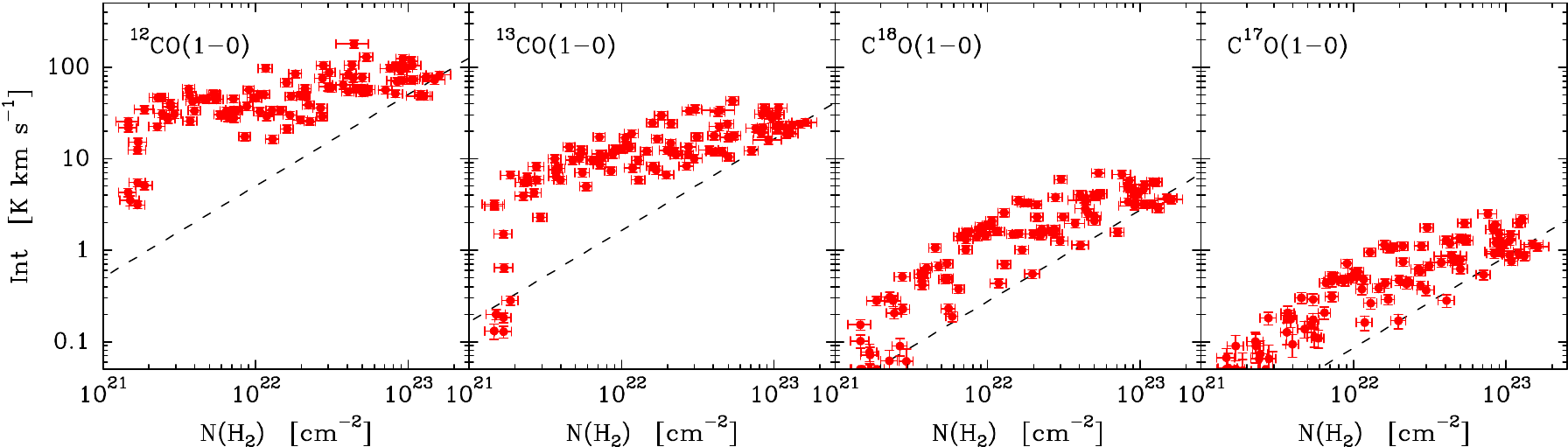}
	\caption{Integrated line intensity as a function of H$_2$ column density for the $J$=1--0
		transition of the four main CO isotopologs. For reference, each panel shows 
		a linear relation that fits the average intensity of the highest column density bin
		(dashed line). 
			Depending on the transition, the lower limit of the intensity scale
			(0.05 K km s$^{-1}$) approximately corresponds to one to three times the noise level
			in the integrated intensity.
	}
	\label{co_3mm}
\end{figure*}

As Fig.~\ref{fig_vs_complete} shows, 
the distribution of the COMPLETE and sampling points agree both in their
correlation with $N$(H$_2$) and in the amount of dispersion for a given $N$(H$_2$),
indicating that the sampling observations recover the general trends of the emission
from the full cloud.
There seems to be a slightly better agreement between the sampling and the COMPLETE
results for the $^{13}$CO(1--0) data, although its cause is unclear
since both CO isotopologs 
were observed simultaneously by the two surveys. We note that small discrepancies between the data are unavoidable given their different calibrations and 
the factor of two difference in angular resolution between
the IRAM 30m telescope used in our sampling survey ($\approx 21''$) and the FCRAO 14m  
telescope used in COMPLETE ($\approx 46''$).

To quantify the comparison between the sampling  
and the COMPLETE results, we calculated for each dataset the intensity
mean and rms inside each of the ten column density bins in which we 
have divided the cloud.
Given the large dispersion of the data, we operated in logarithmic units and
later converted the results to a linear scale.
The results of this calculation are shown in the middle and bottom panels of
Fig.~\ref{fig_vs_complete} in the form of ratios between the estimates derived using
the sampling method and the COMPLETE maps.

As expected from the scatter plots, the sampling/COMPLETE ratio of the means 
(middle panels) is close to unity over the full range of $N$(H$_2$) for both CO isotopologs. The largest deviations from unity occur in $^{12}$CO(1--0), but they
reach at most a factor of about 1.5 and do not show any systematic pattern of bias.
Also as expected from the scatter plots, 
the $^{13}$CO(1--0) ratios are better behaved, and our estimate indicates an agreement 
between the sampling and the COMPLETE mean values at the level of 25\%.
The only disagreement between the $^{13}$CO(1--0) sampling and COMPLETE results
occurs in the lowest column density bin 
($N(\mathrm{H}_2) < 2.2 \times 10^{21}$~cm$^{-2}$), 
where some of the line intensities measured with the sampling technique 
lie below the three sigma detection level of the COMPLETE data (dotted line in the 
scatter plots).
This suggests that the COMPLETE data are too shallow to characterize
the $^{13}$CO(1--0) intensity in the lowest column density bin, and as a result,
they artificially overestimate the mean value and underestimate the dispersion.

Even better agreement between the sampling and the COMPLETE results is found for
the estimate of the emission rms. As shown in the bottom panels of 
Fig.~\ref{fig_vs_complete}, the $^{12}$CO(1--0) data agree better than 25\%, although the
sampling method can barely follow the large rms increase in the lowest column density
bin seen in the upper panel. In hindsight, having observed additional positions in 
this bin would have been desirable.
For $^{13}$CO(1--0), the agreement between sampling and mapping results 
is also better than 25\%, again 
excluding the lowest column density bin due to the insufficient
sensitivity of the shallower COMPLETE data.

To summarize, a comparison with the COMPLETE mapping data 
suggests that the stratified random sampling method can provide estimates of
the intensity mean and dispersion that have an accuracy of better than a factor 
of 1.5, and are usually at the 20\% level, for the whole range of column densities covered
by our survey. This comparison is unfortunately limited to 
CO data since this is the only species for which large-scale maps are available.
While further testing is needed using different species (and clouds), the 
results obtained so far support the idea 
that stratified random sampling
is a potentially useful tool to efficiently 
determine the global properties of the cloud emission.

\subsection{Data overview}
\label{sect_over}

\begin{figure*}
    \centering
    \includegraphics[width=\hsize]{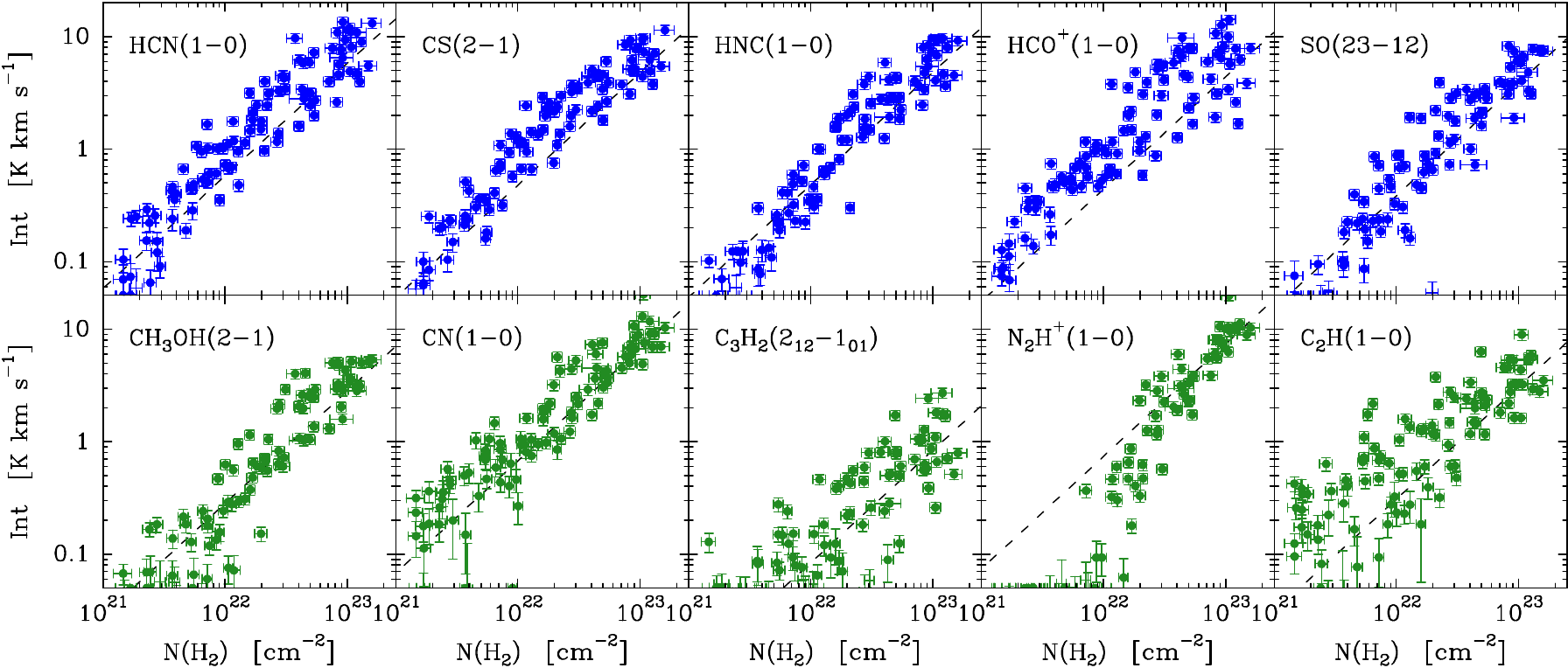}
    \caption{Integrated line intensity as a function of H$_2$ column density for the survey      
    	species detected in at least five column density bins. 
    	{\em Top:} Traditional dense gas tracers  
    	(blue symbols). {\em Bottom:} Additional tracers of dense gas (green symbols). For reference, each
    	panel shows a linear relation that fits the average intensity of the highest column density bin 
    	(dashed line). For HCN(1--0), CN(1--0), N$_2$H$^+$(1--0), and C$_2$H(1--0),
    	the intensity includes the contribution from all detected hyperfine components. For CH$_3$OH(2--1), 
    	the intensity includes the contribution from the A$^+$ and E components.
        	Depending on the transition, the lower limit of the intensity scale
        	(0.05 K km s$^{-1}$) approximately corresponds to one to three times the noise level
        	in the integrated intensity.
}
    \label{dense_3mm}
\end{figure*}

The number of molecular lines detected in each position depends
strongly on its column density. Toward
positions with the highest column densities
(bin number 10, with $N$(H$_2$) $\ge 10^{23}$~cm$^{-2}$), the spectra contain
about 50 different molecular lines 
in the 3mm band alone. As the column density decreases,
the number of detected lines decreases rapidly, and 
in the lowest column density bin
($N$(H$_2$) $\approx 10^{21}$~cm$^{-2}$), the detections
are often limited to the lines of the abundant CO isotopologs.
Since our goal is to study 
the dependence of the line intensity 
with the gas column density, we restrict our analysis 
to those 3mm lines that are detected in most positions belonging to
at least five column density bins out of the ten in which we divided
the cloud. 
In this section, we
focus our discussion on these brighter lines that satisfy our 
selection criterion. 
Table~\ref{tbl_master} presents their integrated intensity 
estimated, as described in Sect.~\ref{sect_obs}, by integrating the
emission over the range of detection, and in case of non detection, 
by integrating the emission over 
the range at which $^{13}$CO(1--0) was detected.
Appendix~\ref{app_stck} illustrates the data presenting 
stacked spectra of all 3mm transitions 
for each column density bin. 

For presentation convenience,
we have divided these lines into three different chemical families.
The first chemical family is that of the CO isotopologs, and consists of
$^{12}$CO, $^{13}$CO, C$^{18}$O, and C$^{17}$O.
Fig.~\ref{co_3mm} shows the velocity-integrated intensity of their
$J$=1--0 transition as a 
function of H$_2$ column density. 
%These intensities were calculated by adding the emission in the spectra as 
%described in Sect.~\ref{sect_obs}.
As shown below, 
the results for the $J$=2--1 line are almost indistinguishable
from those of the $J$=1--0, so we can safely base our analysis on the
low-energy transition.
For reference, each plot contains a dashed line showing
a linear relation that would fit the mean intensity of the
highest column density bin.

The most noticeable trend in Fig.~\ref{co_3mm}
is the strong correlation between all the line intensities and the
H$_2$ column density over the two orders of magnitude that this parameter 
spans across the sample.
Since the sample positions in each column density bin 
are located randomly over the cloud, this correlation 
indicates that the H$_2$ column density
is by itself a strong predictor of the CO line intensity, a first hint that our
reliance on the H$_2$ column density as a proxy for the line emission
is an acceptable choice.

As Fig.~\ref{co_3mm} shows, the brighter $^{12}$CO and $^{13}$CO lines have
the largest dynamic range, and their distribution with H$_2$ column density
presents an abrupt change in slope at around $2\times 10^{21}$~cm$^{-2}$ 
($A_{\mathrm V} \approx 2$~mag). This change has been previously characterized
by \cite{pin08}, who used data from the COMPLETE survey to study the intensity of the
CO isotopologs in Perseus as a function of extinction for  
$A_{\mathrm V} < 10$~mag (equivalent to $N$(H$_2$) $<10^{22}$~cm$^{-2}$).
It likely results from the
photodissociation of the CO molecules in the outer cloud by the UV photons from 
the external interstellar radiation field 
(e.g., \citealt{tie85,van88,leb93,ste95,vis09,wol10,job18}).
Inner to the photodissociation edge, the slope of the $^{12}$CO and $^{13}$CO
intensities is  relatively flat compared with the linear slope, a trend that
will be shown below to result from saturation effects, in agreement with
the previous suggestion from  \cite{pin08}.

For the C$^{18}$O and C$^{17}$O lines, the photodissociation
edge is not appreciable  at low column densities due to insufficient sensitivity, 
although it can be hinted in the bin-averaged data discussed below.
As the column density increases, the C$^{18}$O and C$^{17}$O 
intensities increase almost linearly with $N$(H$_2$) up to about $10^{22}$~cm$^{-2}$,
and then flatten significantly at higher column densities.
This flattening is not caused by optical depth effects since the  
C$^{18}$O(1--0)/C$^{17}$O(1--0) ratio has a close-to-constant 
value of $3.2 \pm 0.1$, which matches the 
$^{18}$O/$^{17}$O isotopic ratio found for the ISM 
by \cite{wil94}, indicating that
both the C$^{18}$O(1--0) and C$^{17}$O(1--0) lines 
are optically thin over the entire Perseus cloud.
As shown by the cloud model discussed below, the flattening is likely 
caused by the freeze out of the CO molecules onto the dust grains 
at the high densities characteristic of the regions with high $N$(H$_2$).

The other two families in which we divide the species detected in our survey are presented in
Fig.~\ref{dense_3mm}. 
The top row of panels shows
species that are commonly used to trace dense gas
both in galactic and extra-galactic 
studies, such as HCN, CS, HNC, HCO$^+$, and SO (e.g., \citealt{eva99,ken12}).
We will refer to these species collectively as ``traditional dense gas tracers,'' with
the caveat that
their role as true dense gas tracers is being reassessed as a result of recent work
(\citealt{kau17}, \citealt{pet17}, \citealt{shi17}, see further discussion in Sect.~\ref{sect_tracers}).
The bottom row contains a more heterogeneous mix of species. Most 
of them are sensitive to dense gas, but they often present
strong sensitivity to additional processes such as
shocks, UV radiation, or CO freeze out:
CH$_3$OH, CN, C$_3$H$_2$, N$_2$H$^+$, and C$_2$H \citep{van98,ber07}.
We will refer to this group of species as the ``additional tracers'' family for lack of a
	better term.

\begin{figure*}
	\centering
	\includegraphics[width=\hsize]{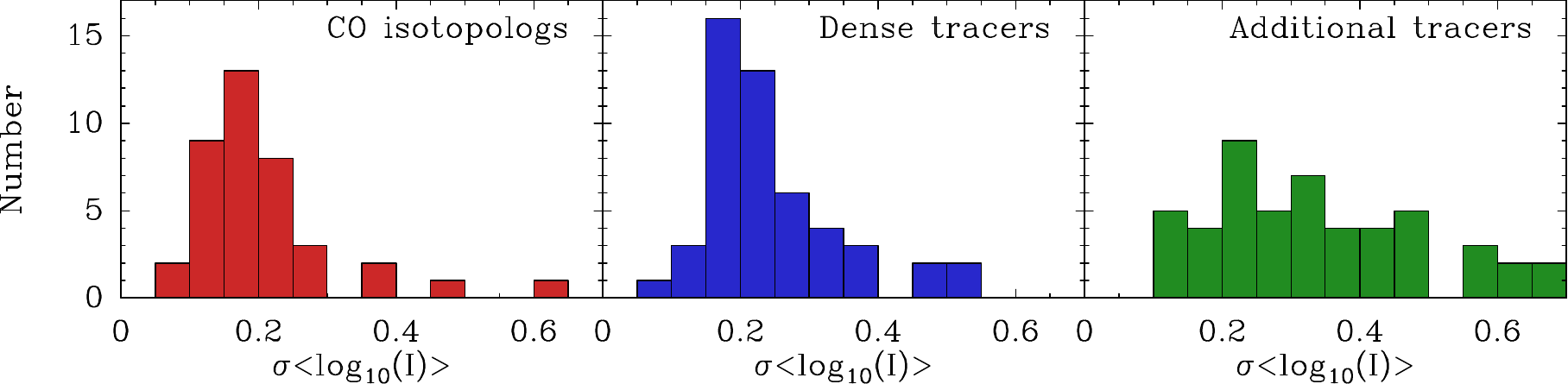}
	\caption{Dispersion histograms of logarithmic intensity for each
		column density bin. The panels show the combined dispersion for the family of CO isotopologs 
		{\em (left)}, the traditional dense gas tracers {\em (middle)}, and the additional tracers  
		{\em (right)}. In all cases, the lowest column density bin has been excluded 
		due to low sensitivity.}
	\label{sigma_hist}
\end{figure*}

As with the CO isotopologs, the intensity of the traditional
dense gas tracers presents a strong correlation
with the H$_2$ column density over the
two orders of magnitude sampled by the survey.
The lower signal-to-noise ratio of these tracers 
at low H$_2$ column density makes it difficult
to judge whether they present a photodissociation
edge similar to that of CO, although there are hints of
such an edge at least in the HCN data.
Better evidence for photodissociation edges in some of these 
tracers will be presented in Sect.~\ref{int_dep} when we analyze the
bin-averaged data.

Inside the photodissociation edge, all traditional dense gas tracers present a
clear quasi-linear correlation with $N$(H$_2$), as indicated
by the good match between the data points and the linear dashed lines.
This systematic behavior of the traditional dense gas tracers
is somewhat unexpected since common  belief among observers suggests
that a linear correlation with column density would only be expected for
the thermalized lines of the thin CO isotopologs. The  traditional dense gas 
tracers, due to their subthermal excitation, are expected to 
trace volume density, not column density, and therefore present a steeper intensity 
slope.
Understanding this unexpected behavior requires a detailed modeling
of the cloud physical and chemical properties, which we defer to 
Sect.~\ref{origin} below. Here  we only stress
that the remarkable linear behavior extends over the two orders of 
magnitude in H$_2$ column density covered by the observations.

The set of additional tracers, shown in the bottom row of Fig.~\ref{dense_3mm}
presents a more diverse behavior in their dependence with the H$_2$
column density.
This behavior ranges from the close to linear
correlation of CH$_3$OH to the 
significantly nonlinear behavior of N$_2$H$^+$,
whose intensity drops precipitously at
H$_2$ column densities below $10^{22}$~cm$^{-2}$. Less
notable, but still significant, are the deviations from linearity
seen in CN and C$_2$H, whose intensities
increase slightly but significantly (factors of two or three) 
over the linear trend for column densities lower than
$10^{22}$~cm$^{-2}$. The possible origin of all these behaviors
is explored below with the help of a radiative-transfer cloud model.

\subsection{Data dispersion}
\label{data_disp}

As mentioned above, the intensity of each line
in the survey correlates strongly with the H$_2$ column density, 
indicating that despite the
random location 
of the observed positions, their line intensity can
be accurately predicted from the value of $N$(H$_2$).
This result is important for the use 
of the stratified random sampling method since, as discussed
in Sect.~\ref{sec-_sampling}, it requires that the
H$_2$ column density is a reliable proxy for the molecular 
emission. To better quantify how well the H$_2$ column density 
predicts the intensity of each line, we have studied the 
dispersion of the intensities inside each column density bin.

For each transition, we calculated  
the mean and rms dispersion of the ten intensity values
belonging to each column density bin.
Since the observations in the 
lowest column density bin often resulted in non detections
(likely due to low
sensitivity and photodissociation effects), we ignored this bin in our
dispersion calculations. In Fig.~\ref{sigma_hist}, we
present histograms of the log-scale rms dispersion
inside each
column density bin for the three chemical families
defined before.

As can be seen, the rms dispersion of the CO isotopologs and the traditional 
dense gas tracers has a well-defined peak near
0.2 dex, a result that does not change 
if we include the data from the lowest column density bins. 
This level of dispersion is equivalent to a factor of 1.6
in linear scale, and its small value is responsible for the tight correlations 
that characterize the intensity-column-density plots.
While low,
the dispersion in the intensities is significantly higher than the dispersion of column
densities inside each bin, which we estimate as 0.05 dex. 
This indicates that the scatter of intensities is intrinsic to the gas, and not
a mere reflection of the range of column densities contained inside each bin.

As Fig.~\ref{sigma_hist} shows, the
family of additional tracers
presents a wider distribution of dispersions and lacks
a clear peak, in agreement with the diversity of 
scatter levels seen in the bottom row of Fig.~\ref{dense_3mm}.
The mean value of this distribution is 0.3~dex,
again independently of whether the lowest column density bin is included.
This value implies that the intensities in this chemical family 
have an rms scatter of a factor of two in linear scale, which is
still much smaller than the two orders
of magnitude spanned by the intensities. This again 
reflects the significant correlation between line
intensities and H$_2$ column densities.

While the low scatter of the intensities
supports the use of the column density as a proxy
for the stratified random sampling,
the fact that the scatter exceeds
what would be expected simply from
column density variations indicates that the column density 
is not a perfect predictor of the emergent intensity.
This should not be surprising given that the column density is
an integrated quantity that can be realized by multiple physical
and chemical conditions along the line of sight.
Our data already suggest several contributions to the
dispersion of intensities associated with a single column density.
Chemical effects, for example, likely play a role in the higher
dispersion seen in the lines of C$_3$H$_2$ and C$_2$H. These two species 
are known to present similar sensitivity to the presence of 
an external radiation field \citep{pet05}, and this field likely
changes significantly across the cloud.
Optical depth effects, in addition, 
are likely to contribute to the dispersion of intensities seen in 
species like HCO$^+$(1--0), whose scatter is significantly larger 
toward the high column density bins. Observations of this line at
selected positions using high velocity resolution reveal that the
HCO$^+$(1--0) lines often suffer from self absorption, a fact confirmed
by the data from the optically
thinner H$^{13}$CO$^+$(1--0), shown in Appendix~\ref{sect_remaining},
which present a much lower degree of scatter.
Another contribution to the scatter in the line intensities 
comes
from differences in the distribution of densities 
(and possible temperatures)
along each line of sight.
This is suggested by an observed increase in the scatter 
of the HCN and CS lines as their level energy increases (Sect.~\ref{sect_model}).
Higher $J$ lines have higher critical densities, and are therefore 
more sensitive to density variations along the line of sight.
While the above examples show the intrinsic limitation of using column density
as the sole predictor of the emergent intensity, they also illustrate how further
understanding of the emission scatter could provide new insights on 
the internal structure of molecular clouds.

\subsection{Correlation with H$_2$ column density}
\label{int_dep}

\begin{figure*}
    \centering
    \includegraphics[width=\hsize]{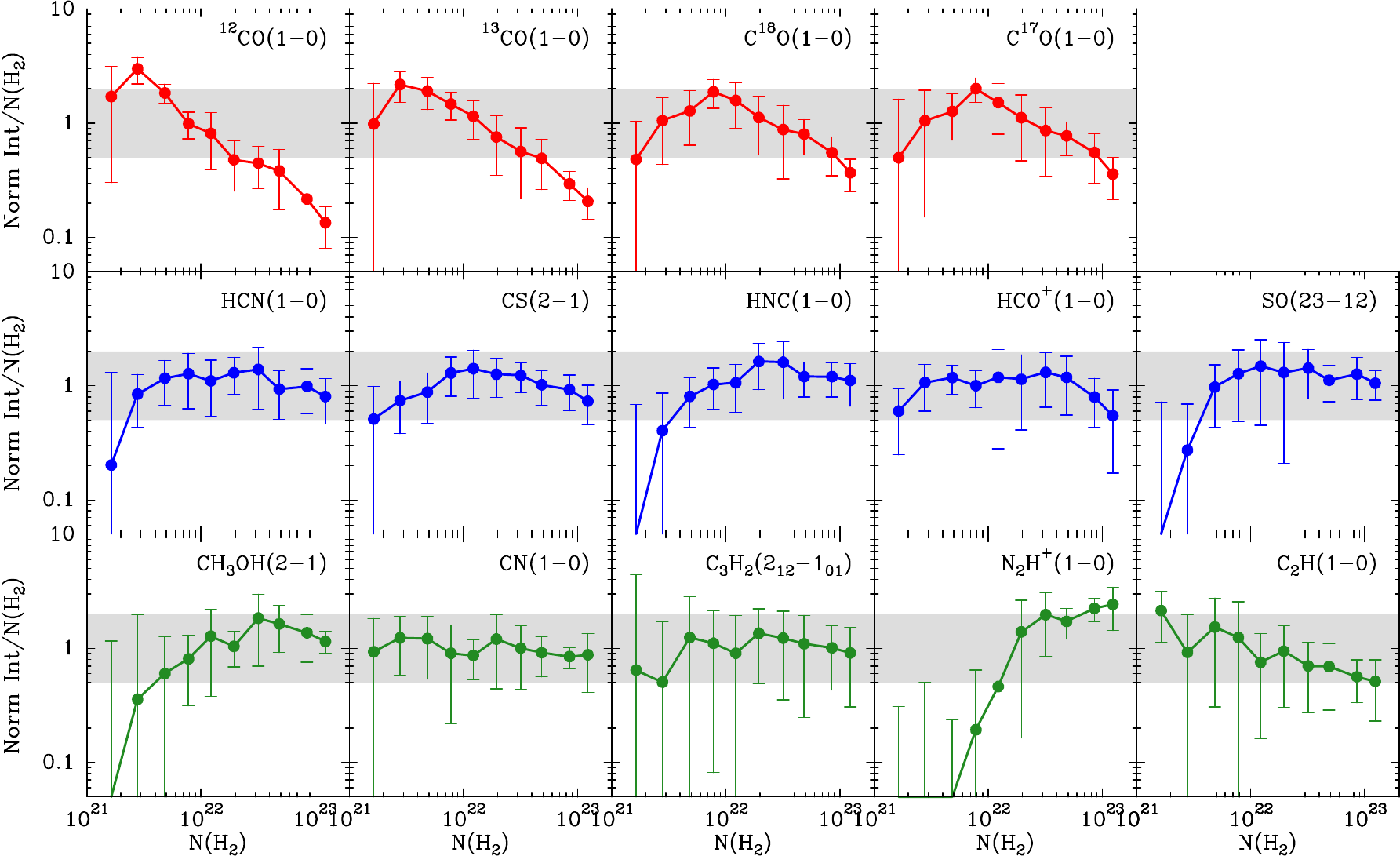}
   \caption{Normalized ratio between integrated intensity and H$_2$ column 
   	density as a function of H$_2$ column density. The data points represent bin-averaged 
   	values and the error bars represent the rms dispersion. The horizontal gray band encloses 
   	the region were deviations from unity are within a factor of 2.}
    \label{linear}
\end{figure*}

\begin{table}
\caption[]{Linear square fits to the $\log$-$Int$ vs. $\log$-$N$(H$_2$) plots.\label{tbl_fit}}
\centering
\begin{tabular}{lcc}
\hline\hline
\noalign{\smallskip}
Transition  & Slope & Pearson-$r$ \\
\hline
\noalign{\smallskip}
\multicolumn{3}{c}{\mbox{CO isotopologs}} \\
\hline
\noalign{\smallskip}
$^{12}$CO($J$=1--0) & $0.37 \pm 0.04$ & 0.70 \\
$^{13}$CO($J$=1--0) & $0.62 \pm 0.06$ & 0.73 \\
C$^{18}$O($J$=1--0) & $0.82 \pm 0.05$ & 0.86 \\
C$^{17}$O($J$=1--0) & $0.71 \pm 0.04$ & 0.87 \\
\hline
\noalign{\smallskip}
\multicolumn{3}{c}{\mbox {Traditional dense gas tracers}} \\
\hline
\noalign{\smallskip}
HCN($J$=1--0) & $0.99\pm 0.04$ & 0.93 \\
CS($J$=2--1) & $1.13 \pm 0.05$ & 0.93 \\
HNC($J$=1--0) & $1.18 \pm 0.04$ & 0.94 \\
HCO$^+$($J$=1--0) & $0.97 \pm 0.05$ & 0.88 \\
SO($N,J$=23--12) & $1.17 \pm 0.06$ & 0.90 \\
\hline
\noalign{\smallskip}
\multicolumn{3}{c}{\mbox{Additional tracers}} \\
\hline
\noalign{\smallskip}
CH$_3$OH($J_k$=2--1) & $1.30 \pm 0.07$ & 0.90 \\
CN($N$=1--0) & $0.90 \pm 0.04$ & 0.93 \\
C$_3$H$_2$($J_{Ka,Kc}$=2$_{12}$--1$_{01}$) & $0.87 \pm 0.07$ & 0.81 \\
N$_2$H$^+$($J$=1--0) & $1.76 \pm 0.08$ & 0.93 \\
C$_2$H($N$=1--0) & $0.78 \pm 0.06$ & 0.78 \\
\hline
\noalign{\smallskip}
\end{tabular}
\end{table}

To quantify how linear the dependence of the intensity with $N$(H$_2$) is, we
made least squares fits to the data points in Figs.~\ref{co_3mm} and \ref{dense_3mm}
(in log-log scale).
To avoid any possible effect of the photodissociation
edge, we excluded the data in the first column
density bin, and for N$_2$H$^+$, we also excluded the
data in the following three bins because no emission was detected in them.
Table~\ref{tbl_fit} presents the fit results (in log-log) together
with the Pearson $r$ coefficient for all the transitions.

As expected from the diagrams, the   
CO isotopologs present slopes that are significantly lower than one,
which is the value that corresponds to a linear correlation.
The lowest slope values are those of $^{12}$CO and $^{13}$CO, 
and reflect the high 
optical depth of these lines, which makes their 
emission only weakly dependent on the gas column density.
The lines from the rare isotopologs C$^{18}$O and
C$^{17}$O are optically thin and present higher slope values.
Still, their lower-than-one slopes indicate that the lack 
of a linear correlation between the intensity and the 
H$_2$ column density represents an intrinsic
property of the CO emission.

Also as expected from the scatter plots, the 
slopes derived for the traditional
dense gas tracers are very close to unity.
The slope values in Table~\ref{tbl_fit} range
between 1.0 to 1.2, and have a scatter at the level
of about 5\%. In addition, the Pearson $r$ coefficients 
of these tracers have similar values 
of about 0.9 indicative of a strong level of correlation.

Most members of the additional tracers group have
slope values similar to those of the traditional dense gas tracers, but a few 
deviate noticeably from a linear behavior.
The most clear case is N$_2$H$^+$, which has a slope
of 1.76 indicative of a preference for high column densities.
More marginal, but probably still significant, is the case of   
CH$_3$OH, which has a slope of 1.30.
At the other end of the scale, C$_2$H presents a significantly
flat slope of 0.78, indicative of a slight intensity increase toward
the low column density gas.

A more graphic representation of the close-to-linear 
relation between some of 
the line intensities with $N$(H$_2$ ) is presented 
in Fig.~\ref{linear}. This figure shows the
ratio between the line intensity and the H$_2$ column
density for each selected transition. For clarity, the plot shows 
normalized ratios with
error bars that indicate the data dispersion inside each bin.

The top panels in Fig.~\ref{linear} (red symbols) show how the CO isotopologs 
significantly deviate from the constant ratio that corresponds 
to a linear correlation between intensity and column density.
This deviation is largest in the optically thick lines of
$^{12}$CO and $^{13}$CO, whose ratio with the H$_2$ column density
varies by more than one order of magnitude over the cloud range.
The thinner
C$^{18}$O and C$^{17}$O lines also present non constant ratios,
but  the variation of their geometrical mean 
over the cloud is limited to a factor
of two up and down, a range that is indicated with a gray-shaded band.
As mentioned above, the emission from these 
two isotopologs is optically thin, so 
the curvature in the plots of Fig.~\ref{linear}
likely arises from systematic variations of the
molecular abundance inside the cloud, an interpretation that
will be further explored with a radiative transfer model 
in Sect.~\ref{sect_model}.

The middle panels in Fig.~\ref{linear} (blue symbols) show
the behavior of the intensity-column density ratio for the traditional
dense gas tracers. Overall, these tracers behave more linearly
than the CO isotopologs, and most of their data points lie inside the
gray-shaded band that encloses variations within a factor of 2.
There is marginal evidence that in most tracers 
the point from the lowest column density bin lays
below the gray-shaded band,  with the possible
exception of HCO$^+$. The most likely cause of this
drop is the photodissociation of molecules by the external UV radiation field.

Finally, the bottom panels in Fig.~\ref{linear} present
the results for the  additional tracers (green symbols).
Most of these tracers have ratios inside the gray-shaded band, with 
the clear exception of N$_2$H$^+$ and possibly CH$_3$OH.
The sudden drop of N$_2$H$^+$ at
H$_2$ column densities lower than
$2 \times 10^{22}$~cm$^{-2}$ is highly significant
since the observations had
enough sensitivity to detect this species at 
low column densities if the intensity had continued
the linear trend. As it will be discussed in
the Sect.~\ref{sect_model}, the drop is consistent with
the N$_2$H$^+$ abundance being only significant
in the denser regions of the cloud where CO is frozen out on the grains,
as previously indicated by observations of dense cores
\citep{cas99, ber02, taf02}.

Although the data points of C$_2$H 
remain in the gray-shaded band of Fig.~\ref{linear},
they present a gradual increase by a factor of four
from high to low column densities.
This increase is again consistent with the expected
abundance enhancement of C$_2$H toward the outer cloud 
caused by the external UV radiation field \citep{pet05}.
The related species CN 
does not present a noticeable increase, although 
the detailed model of the intensities below shows that
it may be slightly enhanced toward the outer cloud.

To conclude the discussion, we provide in Table~\ref{tbl_xfact} 
estimates of the $X$-factor (defined as the 
H$_2$ column density over line intensity)
for each line of the dense-gas and additional tracers as
derived from our least-squares fit. 
For all species except for N$_2$H$^+$, data from all the 
column density bins but the lowest one were used in the fit. 
For N$_2$H$^+$, the fit used data only from the highest four column density bins
since the emission of this species drops non linearly at lower column densities.
Given the approximate linear behavior of all the tracers, the
 $X$-factors in the table could be used to compare the intensity of
 the Perseus emission with that of other clouds. 
 They could also be used to infer H$_2$ column densities 
 from the intensity of the observed lines, although
without a proper calibration using other clouds, the result
will be subject to a great degree of uncertainty.

\begin{table}
\caption[]{$X$-factors for high-dipole moment species.\label{tbl_xfact}}
\centering
\begin{tabular}{lc|lc}
\hline
\noalign{\smallskip}
Transition  & $X$\tablefootmark{a} & Transition  & $X$\tablefootmark{a} \\
% & cm$^{-2}$ (K km s$^{-1}$)$^{-1}$ & & cm$^{-2}$ (K km s$^{-1}$)$^{-1}$ \\
\noalign{\smallskip}
\hline
\noalign{\smallskip}
HCN(1--0) & $1.2 \times  10^{22}$ & CH$_3$OH(2--1) & $3.6 \times 10^{22}$ \\
CS(2--1) &  $1.3 \times 10^{22}$ & CN(1--0) & $1.2 \times 10^{22}$ \\
HNC(1--0) & $2.1 \times 10^{22}$ & C$_3$H$_2$(2$_{12}$--1$_{01}$) & $1.1 \times 10^{23}$ \\
HCO$^+$(1--0) & $1.2 \times 10^{22}$ & N$_2$H$^+$(1--0) & $1.4 \times 10^{22}$ \\
SO(23--12) & $2.5 \times 10^{22}$ & C$_2$H(1--0) & $2.2 \times 10^{22}$ \\
\hline
\noalign{\smallskip}
\end{tabular}
\tablefoot{
\tablefoottext{a}{$X = N(\mathrm{H}_2) /$ Intensity, in units of cm$^{-2}$ (K km s$^{-1}$)$^{-1}.$}
}
% for all trans, last bin not used. For N2H+, only 4 bins used
\end{table}

\subsection{Comparison with other clouds}
\label{sect_compare}

As mentioned in the Introduction, a recent effort has been made by
different authors to 
map the emission of entire or close-to-entire molecular clouds in multiple lines.
In this section we compare the results from this effort to our
observations, both to test the sampling technique and to study
the behavior of the emission in different molecular clouds.

We first compare our dataset with that of 
\cite{wat17}, who carried out a multiline study of
the high-mass star-forming cloud W51. These authors found that
the 3mm-wavelength emission from W51 is 
dominated by the lines of the CO isotopologs together with
the same  traditional dense gas tracers found by us in Perseus (HCN, HCO$^+$, HNC, and CS).
Lacking extinction measurements for W51, 
\cite{wat17} used the integrated intensity of $^{13}$CO(1--0)
as a proxy for the cloud column density, and found a
tight and often close-to-linear correlation between this tracer and 
the integrated intensity of most molecular lines.
This result is similar to our finding of a tight correlation 
between the intensity of the
main molecular species and the H$_2$ column density in Perseus,
with the caveat that in Perseus the $^{13}$CO(1--0)
emission does not correlate linearly with
the H$_2$ column density (although in contrast with Perseus, 
the $^{13}$CO(1--0) emission in W51 is optically thin, see \citealt{wat17}).

A better comparison to our work can be made with the 
large-scale mapping of Orion B by \cite{pet17}.
These authors used extinction measurements from \cite{lom14} to
study the correlation between the intensity of the
different lines and the extinction, as we have done with the Perseus data.
In good agreement with our results, these authors
find a tight correlation that is often close to linear for 
column densities larger than a threshold value equivalent to an extinction of 
$A_\mathrm{V}$=1-3~mag (their Fig.~10). 

Further insight on the Orion B molecular emission 
comes from the  principal component analysis
of \cite{gra17} using the same dataset as \cite{pet17}.
This analysis shows again that
the main predictor of the cloud emission is the gas column density, 
which is responsible 
for the first principal component (60~\% contribution).
The following principal components (at the 10~\% level) reveal
several species whose emission presents a peculiar behavior:
N$_2$H$^+$ and CH$_3$OH, classified as sensitive to gas density, 
and C$_2$H and CN, classified as
sensitive to UV. These species are the same as those identified as peculiar
in Perseus (Sect.~\ref{int_dep}), and this common behavior 
suggests that despite their very different characteristics 
(Orion B contains several embedded HII regions and the bright Horsehead PDR),
the molecular emission from these two clouds 
is controlled by the same main mechanisms.

The final dataset with which we compare our Perseus results is that presented
by \cite{kau17} for Orion A, which is more limited than our Perseus dataset in number of
transitions. \cite{kau17} focus their study of dense gas tracers on 
HCN, CN, C$_2$H, and  N$_2$H$^+$, which are four of the 11 species we studied
in Perseus. For the first three species, they
find that the ratio of the integrated intensity over the gas column density
varies little over the cloud, and
decreases at most by a factor of two when the column
density increases by one order of magnitude
(their Fig.~2). This behavior is similar
to the almost constant intensity-column density ratio
found by us in Perseus. 

Also in agreement with our Perseus results, the N$_2$H$^+$ emission from Orion A 
differs from the other tracers by increasing rapidly in
the most opaque gas (with the possible 
exception of the densest two bins).
As \cite{kau17} discuss, the N$_2$H$^+$ emission seems to be the only 
tracer that is sensitive to the densest component of the cloud.

While limited in targets, the above studies and our Perseus data
cover a variety of clouds with different levels of star-formation activity
both in the low and high-mass regimes.
Despite these differences, all clouds 
present systematic and often tight correlations 
between the emission from most molecular tracers
and the gas H$_2$ column density. 
This correlation indicates that the cloud H$_2$ column density 
behaves as a reliable proxy 
of the molecular emission under a variety of cloud conditions, 
which we saw was a requirement for the stratified sampling
method to work. 
The data, therefore, support the idea that 
the stratified sampling method could be used to characterize the
emission from a large variety of clouds.

The data also show that there are significant similarities between the emission from the 
different clouds, both in terms of the brightest lines and the dependence 
of their intensity with the H$_2$ column density.
Clearly more work needs to be done in this
area, especially by comparing clouds using the same 
observing technique. The initial results, however, suggest that despite the large
differences between the clouds in terms of mass and star-formation activity,
the emission that they produce follows a relatively simple and similar pattern.

\section{A simple emission model for the Perseus cloud}
\label{sect_model}

\begin{figure}
   \centering
   \includegraphics[width=\hsize]{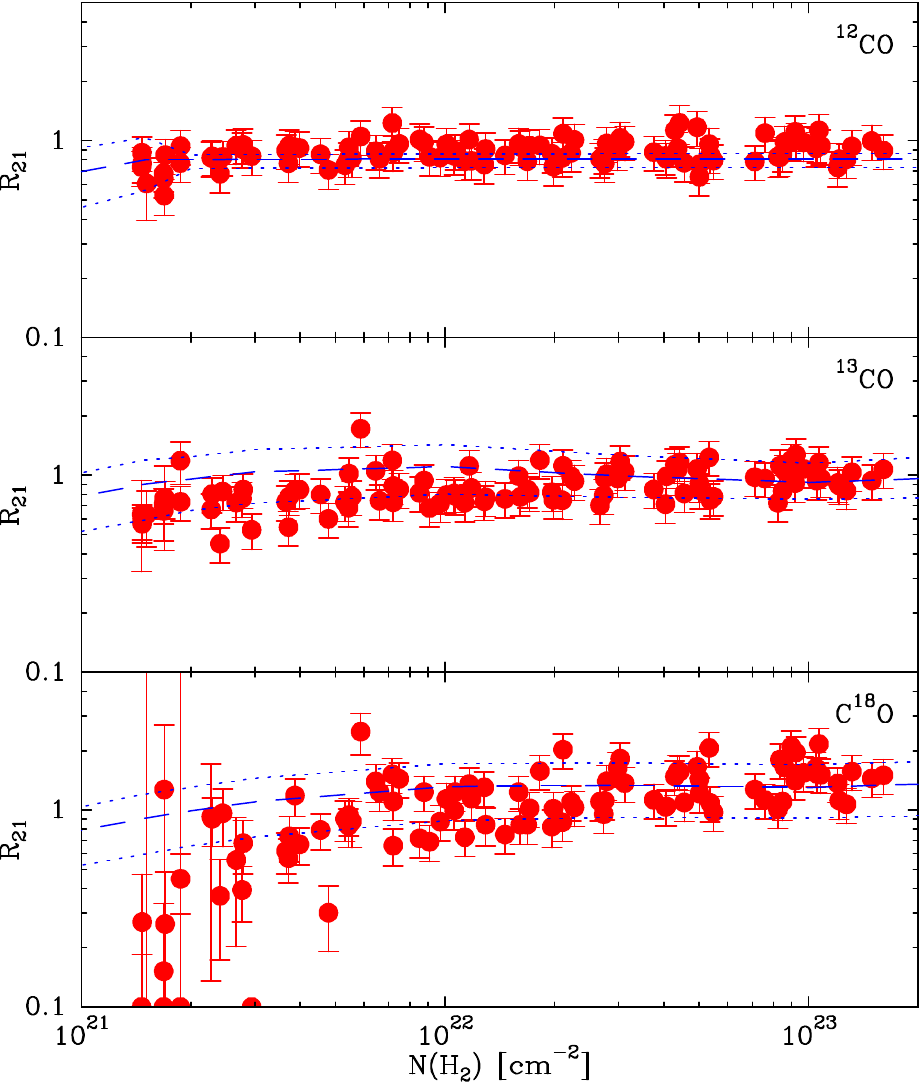}
   \caption{Ratio between $J$=2--1 and $J$=1--0 integrated intensities as a function of H$_2$ 
   	column density for the three CO isotopologs for which both transitions were observed 
   	(red circles). The blue dashed line
   	represents the prediction from a radiative transfer model of an isothermal cloud at 
   	11~K. The dotted lines represent similar models for 15~K (higher 
   	curve) and 7.5~K (lower curve).}
   \label{ratio_21}
\end{figure}

The systematic and relatively simple behavior of the line intensities
in Perseus suggest that the emission from the
cloud must be controlled by the global properties of its gas,
and not by specific details of different parts of the cloud.
If this is the case, it should be possible to 
reproduce the emission behavior using a relatively simple 
treatment of the gas physical and chemical properties.
In this section, we explore this possibility by building a
simple cloud model that 
reproduces simultaneously the main emission properties 
identified 
by our Perseus survey.
While this model results from a deliberate attempt
to reproduce the 
observations, it should be not be seen as the product of 
a systematic search for
the best possible fit, but as an illustration of how
the intensities observed in Perseus 
can be naturally explained as arising from
gas conditions expected to occur in a molecular cloud
of its type.

\subsection{Physical parameters}
\label{sect_phys}

Since our data show that the Perseus line intensities depend 
to first order on
the H$_2$ column density, we modeled the cloud using the
H$_2$ column density as the main physical parameter.
All the other gas properties that contribute to the line intensity,
such as the temperature, density, velocity dispersion, and 
molecular abundances were
modeled as functions of the cloud H$_2$ column density
using simple parametric expressions.
The form of these expressions was determined by 
fitting the intensity of transitions known to be sensitive to specific
physical properties. Table.~\ref{tbl_model_ph} summarizes 
our choice for the physical parameters, and Appendix~\ref{app_abu} 
does the same for the molecular abundances.

To determine the cloud gas temperature profile,
we used the 2--1/1--0 intensity ratio of $^{12}$CO, $^{13}$CO, and C$^{18}$O
since these species span a large range of optical depths and are 
easily thermalized due to their low dipole 
moment.
As Fig.~\ref{ratio_21} shows, the 2--1/1--0 ratio of the three CO isotopologs 
is approximately
constant over the full range of column densities, suggesting that an isothermal 
solution may be able to fit the data.
Such type of solution would also be in line with the estimated dust 
temperature, which presents an approximately constant 
value over the whole cloud \citep{zar16}.
A natural choice for the model gas temperature is 11~K since it corresponds
to the average value determined by \cite{ros08a} using NH$_3$ observations of almost 200 
dense cores.
Fig.~\ref{ratio_21} shows that 
11~K indeed provides a reasonable fit to the data (blue dashed lines), and while 
the fit quality could be slightly improved by adding
small ad hoc deviations from isothermality,
the constant temperature profile was preferred in the name of 
simplicity. \footnote{We note that our analysis does not correct
for differences in the angular resolution
of the J=2--1 and 1--0 observations since we cannot predict how
the beam mismatch will affect the line intensities. Applying
the standard beam dilution correction, for example, would be inappropriate because this 
correction
assumes that the emission arises from a small source located at the beam center, while our
observations deal with extended emission observed with a random sampling.
Not applying a beam correction will likely increase the noise in the line ratio, but
will avoid introducing a systematic bias.
The good behavior in Fig.~\ref{ratio_21} of the $^{12}$CO line ratio,
which is temperature insensitive and therefore cannot be compensated with a special
temperature choice, seems to support our approach.}

A constant temperature solution may at first seem surprising since the cloud outer 
layers are likely warmer than the interior \citep{wan83}. Detailed modeling of molecular 
cloud surfaces by \cite{wol10}, however, 
shows that the CO-emitting gas that we used for the temperature
determination is almost insensitive to this
outer warming because the same UV photons that are responsible for the
warming are also responsible for photodissociating CO (see also \citealt{glo12}).
The net result of this process is that 
most of the warm outer gas is CO dark, and most of the CO-emitting gas remains 
at a close-to-constant temperature of 10~K. This effect can be best seen in Figs.~4 and 6
from \cite{wol10},
which show that the gas temperature remains close to 10 K
all the way from the cloud interior to an $A_{\mathrm V}$ of about 2,
which corresponds to the inner edge of the lowest column density bin in our sampling.
Only the outer 1 $A_{\mathrm V}$ magnitude of the cloud contains CO-emitting gas that is warmer
than 10~K, a fact that our modeling cannot test well due to the weak
signal of the emission. Any temperature increase in the outer cloud will
therefore only affect our modeling of the already poorly constrained outermost bin.

The second cloud parameter that we model is the volume density of the gas. Assigning a single
volume density to a given column density represents a very strong 
simplification since any line of sight through the cloud 
likely contains densities that vary by more than one
order of magnitude. As discussed below, this simplification limits the quality of the fits,
but unfortunately is necessary if we are to use a simple radiative transfer model.
To determine the best-fit density profile we used a more indirect
approach than for the temperature since no molecular
line or line ratio is significantly more sensitive to this
density than others.
After some experimentation, we chose to fit 
simultaneously the emission from multiple transitions of
the traditional dense gas tracers HCN (J=1--0 and 3--2) and CS (J=2--1, 3--2,
and 5--4) because they approximately span the range of upper level energies
covered by our line survey.
The results of this fitting are shown below since they also depend on the
abundance profiles discussed in Sect.~\ref{sect_abu}.
Here we just state that the fit requires a gas volume
density that depends on the column density approximately 
as a power law of the form  $N$(H$_2$)$^{0.75}$. 

While the above density profile represents our favored choice to fit the observed 
line emission,
it should be considered only as a model parameterization. It 
represents a not-well-defined 
line-of-sight average weighted by the emissivity of the different    
lines, and it likely spans a limited range of values
compared with the true range of volume densities in the cloud.
This can be seen from the fact that if we were to use
a similar relation to determine the spatial extent of the gas in the different 
density regimes assuming simply that $N({\mathrm H}_2) = n({\mathrm H}_2)  \times L$,
a steeper density profile would be required to reproduce the observed larger extent of the
lower-density gas. Unfortunately, no similar global fit of the emission 
has been carried on other clouds,
so it is not possible to compare our results with previous work.
We note however, that a similar (or close to linear) density 
dependence with column density can be seen in the compilation of 
numerical simulations presented by \cite{bis19}  (their appendix B).

\begin{table}
	\caption[]{Model physical parameters \label{tbl_model_ph}}
	\centering
	\begin{tabular}{lc}
		\hline
		\noalign{\smallskip}
		Parameter & Value  \\
		\noalign{\smallskip}
		\hline
		\noalign{\smallskip}
	    $T_{\mathrm K}$  &  $11$~K \\
		$n(\mathrm{H}_2)$ &  $2 \times 10^4\; \mathrm{cm}^{-3}\; 
		(N(\mathrm{H}_2)/10^{22}\; \mathrm{cm}^{-2})^{0.75}$ \\
		$\Delta V$\tablefootmark{a} & $1\; \mathrm{km s}^{-1}\; 
		(N(\mathrm{H}_2)/10^{22}\; \mathrm{cm}^{-2})^{0.15}$ \\
		\hline
	\end{tabular}
\tablefoot{
\tablefoottext{a}{Linewidths of $^{12}$CO and $^{13}$CO were multiplied by additional 
	factors of 4 and 1.75 respectively to match observations. See text and Fig.~\ref{linewidth}.}
}
\end{table}

\begin{figure}
	\centering
	\includegraphics[width=\hsize]{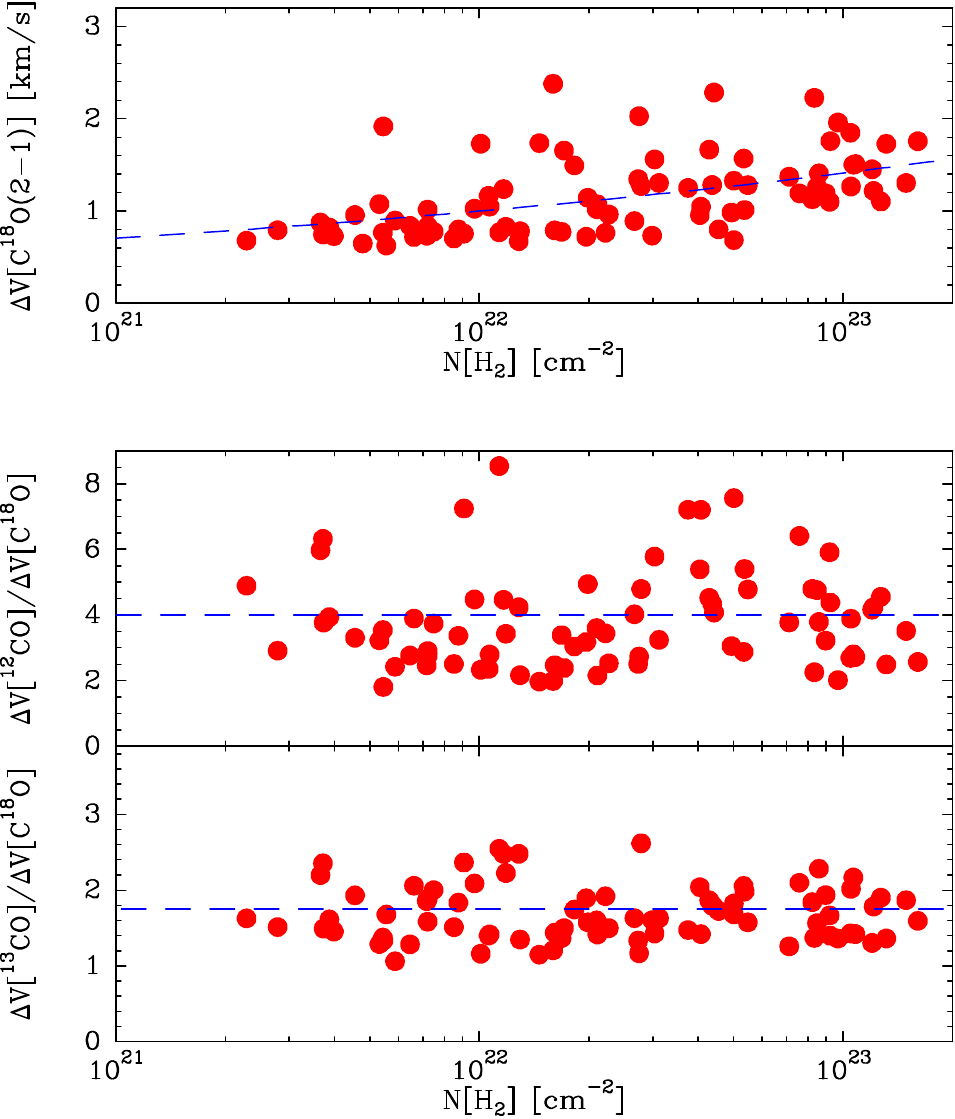}
	\caption{
		Gas internal motions.
		{\em Top:} C$^{18}$O(2--1) FWHM linewidth as a function of H$_2$ column density derived
		from Gaussian fits to the spectra (red circles). The dashed line
		represents an analytic approximation used to model the trend. 
		{\em Bottom:} Linewidth ratios of the two main CO isotopologs
		over C$^{18}$O. The dashed lines represent the constant ratios used in the model (4 for
		$^{12}$CO and 1.75 for $^{13}$CO).}
	\label{linewidth}
\end{figure}

The final physical quantity required by our model is the gas velocity dispersion.
We parameterized it using the observed linewidth of
C$^{18}$O(2--1) since this line is optically thin and
was observed with relatively 
high velocity resolution (0.27 km s$^{-1}$) due to its higher frequency.
The top panel of Fig.~\ref{linewidth} shows that the C$^{18}$O(2--1)
linewidth increases weakly with 
H$_2$ column density in a way that we parameterized as 
$N$(H$_2$)$^{0.15}$ (dashed line). 
This linewidth increase with column density 
is likely caused by the star formation activity in the 
high column density bins. As Fig.~\ref{map} shows, these bins
are concentrated in the main star-forming regions of the cloud
(NGC 1333, B1, and L1448), and many of their CO spectra present wings
indicative of outflow contamination.
Surveys of both Taurus and Perseus have shown that the linewidth of the C$^{18}$O
lines tends to be larger in dense cores with stars compared to starless cores
\citep{zho94,kir17}, and the survey of Taurus cores by \cite{oni96} found a systematic
increase in the C$^{18}$O linewidth with H$_2$ column density similar to the one found
by us in Perseus.

The lower panels of Fig.~\ref{linewidth} show
that the J=2--1 linewidth of the more abundant isotopologs
$^{13}$CO and $^{12}$CO is significantly larger than that of C$^{18}$O(2--1),
by factors of 1.75 and 4, respectively.
These larger linewidths most likely result from optical depth broadening
(e.g., \citealt{hac16}), and since the radiative transfer model 
described below does not reproduce this feature,
we incorporated them explicitly into the model
when predicting the $^{13}$CO and $^{12}$CO emission.

\subsection{Chemical abundances}
\label{sect_abu}

\begin{figure*}
	\centering
	\includegraphics[width=\hsize]{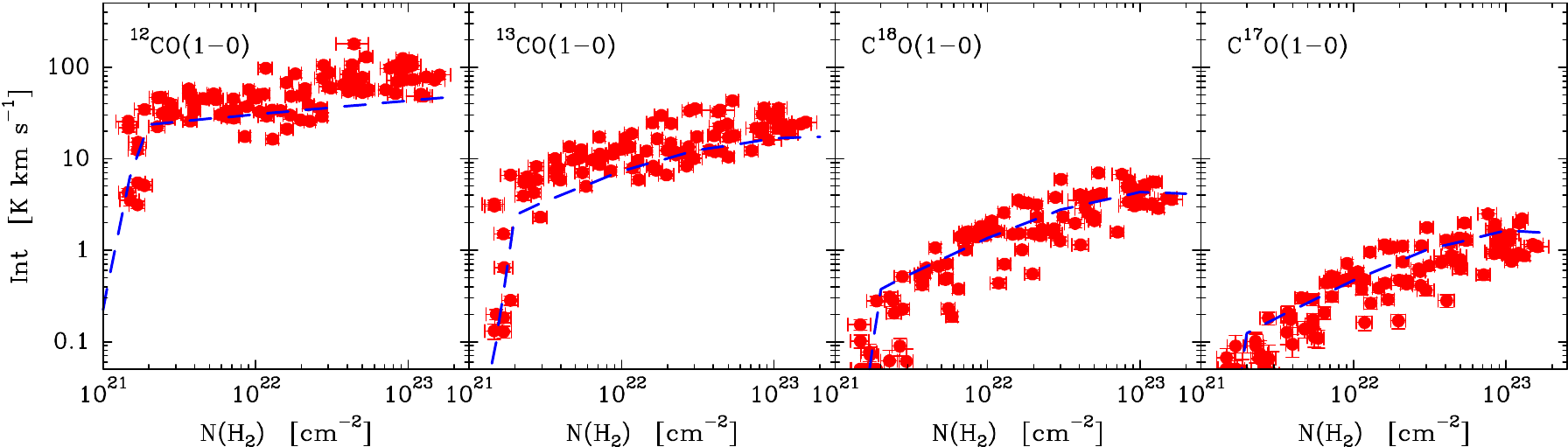}
	\caption{Comparison between observations and model results 
		for the family of CO isotopologs. 
		The red symbols represent the observed 
		intensities of the $J$=1--0 line, previously presented in Fig.~\ref{co_3mm}, and
		the dashed blue lines represent the model results. The sharp
		drop at low column densities is caused by the photodissociation edge in the abundance profile.}
	\label{lvg_co}
\end{figure*}

Once the physical properties of the model cloud were fixed, 
the only parameter left to fit the emission of each species was its 
abundance profile.
We aimed to describe these profiles using a parameterization
that is both simple and consistent with our
current knowledge of the chemical processes 
occurring in a molecular cloud like Perseus.
After some exploring, we found that reasonable fits can be obtained by
using a parameterization containing three terms:
\begin{equation}
\label{equ_abu}
X({N(\mathrm{H_2}})) = X_0 \times f_{\mathrm{out}}(N({\mathrm{H_2}})) 
\times f_{\mathrm{in}}(N({\mathrm{H_2}})),
\end{equation}
where $X_0$ represents a constant scaling factor, 
and $f_{\mathrm{out}}$ and $f_{\mathrm{in}}$
are two normalized factors that represent abundance changes with respect
to $X_0$ in the outer and inner parts of the cloud (i.e., at low and high
H$_2$ column densities).

Appendix~\ref{app_abu} provides a detailed discussion of the
meaning of the different factors and the analytic expressions 
used in the model.
In this section, we summarize the main ideas that are required to understand 
the model results presented below.

The $ f_{\mathrm{out}}$ factor represents any
abundance change that takes place in the outer layers of the cloud,
likely as a result of their exposure to the external UV radiation
field. It is dominated by the
contribution of molecular photodissociation in a thin outer layer
of a few magnitudes of extinction, a process that has been modeled in 
great detail by previous chemical work (e.g., \citealt{tie85,van88,leb93,ste95,vis09,wol10,job18}).
In our model, a simple photodissociation drop seems enough 
to fit the observed intensities of most species in the lowest 
column density bins. For these species, we used an
analytic formula based on the realistic PDR models of \cite{rol07}.
This formula is presented in Appendix~\ref{app_abu}, and
has as its only free parameter the location of the sharp edge 
expressed in units of $A_\mathrm{V}$. 
Changes in this parameter allow the model to adjust for the still
poorly characterized value of the UV radiation field in the cloud, which
has been previously estimated 
to have a Draine $\chi$ parameter \citep{dra78})
between 1-3 \citep{pin08} or 24 \citep{nav19}.
It should be noted, however, that
the low column-density intensities for most species are too weak
to constrain the location of the edge, so most lines were fitted with a 
fixed value of $A_\mathrm{V} = 2$~mag (the value derived for CO).

For C$_2$H and CN, the data show an intensity enhancement
in a layer interior to the photodissociation edge, so
we complemented the drop term with a more gradual outward 
abundance enhancement. This term is based on the detailed modeling 
of the Orion Bar PDR by \cite{cua15}, who found evidence for
an outer abundance increase in the
small hydrocarbons driven by gas-phase reactions involving C$^+$.

The final factor in our abundance parameterization 
is $f_{\mathrm{in}}$, which describes possible variations
in the cloud interior (i.e., at high $N$(H$_2$) values). For all species except N$_2$H$^+$, the
data require a systematic abundance decrease with $N$(H$_2$), 
likely caused by freeze out. This
is consistent with previous findings toward starless dense cores in different environments
\citep{cas99,ber02,taf02}, and for this reason, we parameterized
$f_{\mathrm{in}}$ with an expression used by \cite{taf02}
to describe such systems. This expression
has as only free parameter the volume density characteristic of 
freeze out, which has been adjusted for each molecular species. 
As with the photodissociation edge, a narrow range of choices
(1-$2 \times 10^5$~cm$^{-3}$) is enough to fit all the observations.

For N$_2$H$^+$, the observations require an abundance 
enhancement toward the cloud interior, in agreement with the theoretical
expectation that this species increases its abundance when CO freezes out
\citep{ber97, aik01}. To parameterize this effect, we used a simple
expression related to that used for freeze out,
and which is further described in Appendix~\ref{app_abu}.

As an additional constraint to the model, we required that the
relative abundances of the isotopologs follow the 
ratios determined for the local ISM by \cite{wil94}.
We thus assumed the following isotopic ratios: 
77 for $^{12}$C/$^{13}$C, 560 for $^{16}$O/$^{18}$O, 
3.2 for $^{18}$O/$^{17}$O, and 22 for $^{32}$S/$^{34}$S.

\subsection{Model results}
\label{sect_mod_resu}

\begin{figure*}
	\centering
	\includegraphics[width=\hsize]{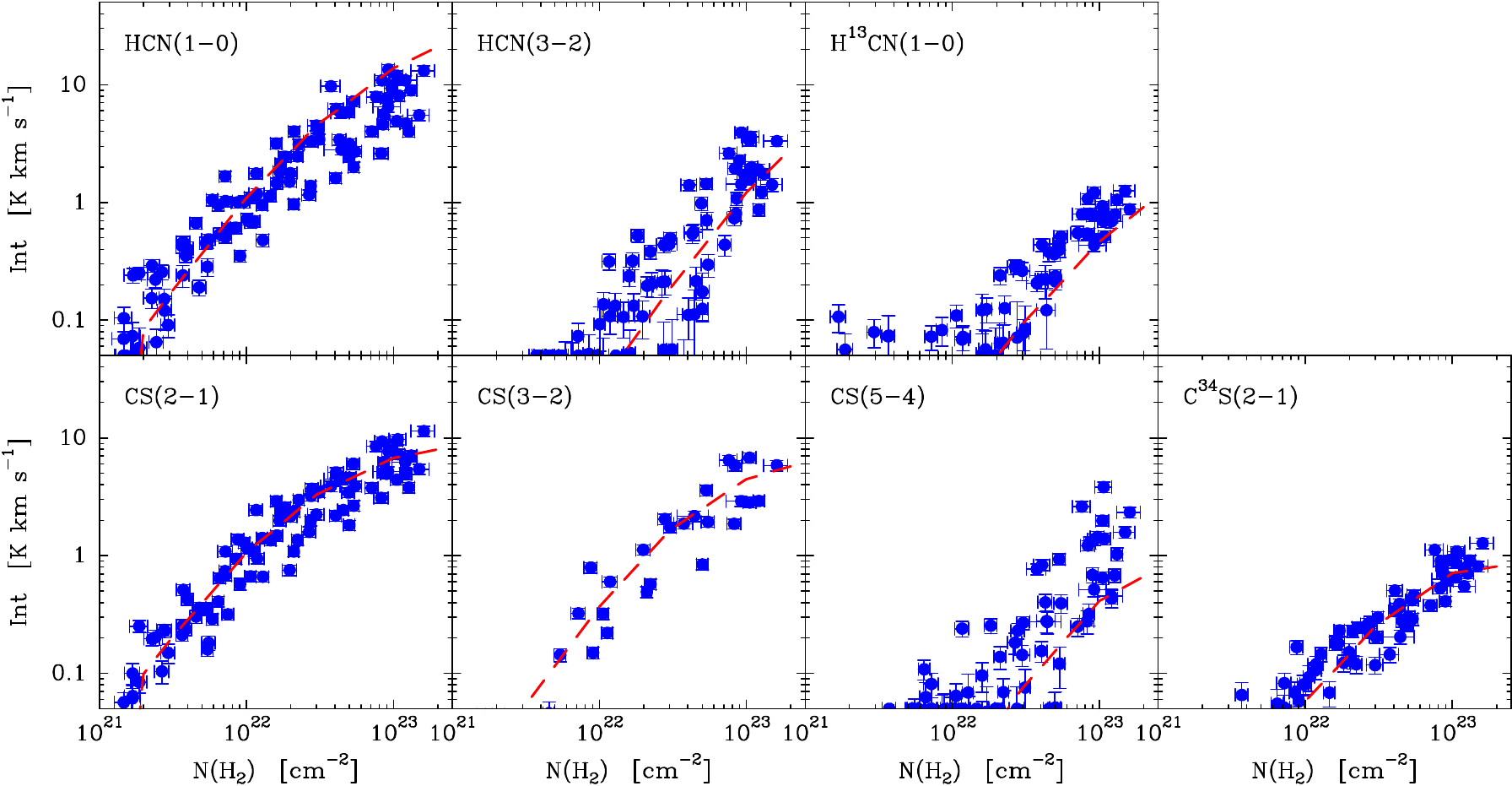}
	\caption{Comparison between observations and 
		model results for HCN {\em (top)} 
		and CS {\em (bottom)}.
		Each row presents the intensity of the different transitions observed for the species 
		(main and rare isotopologs)
		as a function of H$_2$ column density (blue symbols) together with the predictions
		from the model (dashed red lines). 
		The CS(3--2) dataset has fewer points due to the limited observations carried out in the 
		2mm-wavelength band. Several H$^{13}$CN(1--0) points can
		be seen above the
		plot lower limit at very low column densities. A visual 
		inspection of their spectra suggests that they represent 
		noise or baseline residuals and not true molecular emission.
	}
	\label{lvg_hcn_cs}
\end{figure*}

To predict the emergent line intensities from the cloud model,
we used a large velocity gradient (LVG) approximation to the
radiative transfer \citep{cas70, sco74}. This approximation
provides a reasonable estimate of the emission 
given the uncertain geometry of the cloud \citep{whi77},
and thanks to its speed, allows exploring efficiently a large 
range of input cloud parameters.
Our LVG code is based on that presented by \cite{bie93},
and was complemented with
molecular data compiled by the 
Leiden Atomic and Molecular Database 
(LAMDA),\footnote{\url{https://home.strw.leidenuniv.nl/~moldata/}}
which is continually updated with the most recent literature values
\citep{sch05,van20}.
In this section, we present the results of modeling 
several representative species that illustrate the different emission
behaviors identified in the cloud. Modeling results
for the rest of the species 
and a table summarizing the abundance 
parameters derived from the fits 
are presented in Appendix~\ref{sect_remaining}.

Fig.~\ref{lvg_co} shows the model results for the $J$=1--0
transition of the CO isotopologs
(dashed blue lines). Similar results were obtained 
for the $J$=2--1 transitions, as can be inferred from the 
fit to the 2--1/1--0 ratios shown in Fig.~\ref{ratio_21}.
The $X_0$ value of the different CO isotopologs was set 
by fixing the value of C$^{18}$O to the
determination by \cite{fre82} ($= 1.7\times 10^{-7}$) and using the
already-mentioned isotopic ratios from \cite{wil94}.
In addition, the model assumed a photodissociation
edge of $A_{\mathrm V}$ = 2~mag and a freeze-out critical density
of $10^5$~cm$^{-3}$ for all CO isotopologs.

As can be seen from Fig.~\ref{lvg_co}, the cloud model, although not perfect,  
fits reasonably well
the emission. 
This is remarkable 
given the simplicity of the model and the
small number of free parameters used to adjust the fit since
once the cloud physical parameters have been fixed, only 
three free parameters ($X_0$, the $A_\mathrm{V}$ value of the 
photodissociation edge, and the freeze-out density) are left to
reproduce the intensity of the four CO isotopologs over two orders
of magnitude in H$_2$ column density.
According to the model, the $^{12}$CO lines are
optically thick and thermalized everywhere inside the photodissociation 
edge, and the slight intensity increase with the H$_2$
column density arises from the similar increase in the linewidth.
The $^{13}$CO line, on the other hand, only becomes 
optically thick 
at H$_2$ column densities higher than $2 \times 10^{22}$~cm$^{-2}$.
As expected, the C$^{18}$O and C$^{17}$O lines are optically 
thin everywhere inside the cloud.

A number of shortcomings in the model can be attributed
to our simplified treatment of the abundance of the different
CO isotopologs. Our model assumes that they are scaled versions of each other, while 
in reality the $^{13}$CO abundance is expected to increase with respect to 
$^{12}$CO near the cloud edge due to isotopic fractionation,
and the C$^{18}$O and C$^{17}$O abundances are expected to decrease
in the same region due to isotope-selective
photodissociation \citep{bal82,van88}. These two effects would likely enhance
the $^{13}$CO intensity and decrease the C$^{18}$O and C$^{17}$O 
intensities in the vicinity of the photodissociation edge,
improving the fit results.

\begin{figure}
   \centering
   \includegraphics[width=0.8\hsize]{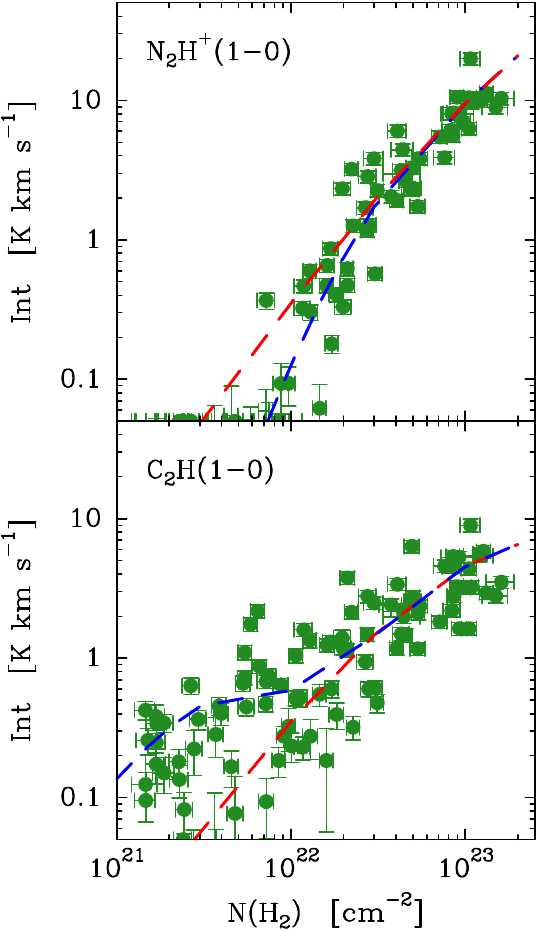}
   \caption{Comparison between observations and 
   	model results for two species that require special 
   	abundance profiles: N$_2$H$^+$ {\em (top)} and C$_2$H {\em (bottom)}.
   	In both panels, the green symbols represent the observed intensity,
   	previously shown in Fig.~\ref{dense_3mm}.
   	In the N$_2$H$^+$ panel, the dashed red line represents 
   	a constant abundance model that fits the intensity at high column densities.
   	The blue dashed line represents the best-fit model, and includes an internal abundance enhancement.
   	In the C$_2$H panel, the dashed red line represents a model with no external enhancement, 
   	while the blue line represents the externally enhanced best-fit model.}
   \label{lvg_n2hp_c2h}
\end{figure}

To explore the ability of the cloud model to fit the intensity of the
traditional
dense gas tracers, we focus here on HCN and CS, which are the most
widely used members of this family. Appendix~\ref{sect_remaining}
presents the results for the remaining members.
For HCN and CS, our observations provided intensities for all sample positions
of the  transitions in the 1 and 3mm wavelength bands, and for CS, a limited
number of positions were also observed in the 2mm wavelength band. 
In addition, the 3mm transitions of
H$^{13}$CN and C$^{34}$S were also detected and included in the model.

Fig.~\ref{lvg_hcn_cs} compares all the available HCN and CS data 
(blue symbols) with the predictions from our cloud model (dashed red lines).
The model assumes
the same abundance factor $X_0$ ($=3\times 10^{-9}$) and
photodissociation edge ($=2$~mag) for both species, 
while the freeze-out critical density
is $2\times 10^5$~cm$^{-3}$ for HCN and $10^5$~cm$^{-3}$ for CS.
As with the CO isotopologs, a simple three-parameter model
 approximately fits simultaneously
all the observed line intensities over the full range of cloud column densities.

While the model intensities remain within the scatter of the data points,
and are therefore consistent with the observations, Fig.~\ref{lvg_hcn_cs}
shows that the model does not provide the best possible fit.
The HCN model
slightly overestimates the $J$=1--0 intensity 
of the main isotopolog while it underestimates the
intensity of H$^{13}$CN.
For CS, the model reproduces well the $J$=2--1 transition of
both the main and rare isotopologs together with the 3--2 intensity,
but it is close to the lower boundary of the $J$=5--4 data points.
Both deviations, and similar ones found in the modeling of HNC and HCO$^+$
(Appendix~\ref{sect_remaining}), likely result from
the use of a single density value to represent the complex cloud structure
along any given line of sight.  
In a real cloud, the density along any line of sight likely increases
toward the interior. As a result, an optically thick line
like HCN(1--0) will sample lower densities than an optically
thin line like H$^{13}$CN(1--0). This effect 
will decrease the HCN(1--0) intensity and 
increase the H$^{13}$CN(1--0) intensity compared to our model, in agreement with 
the observations. Also, a high critical density line, like CS(5--4), will be
sensitive to the highest density gas along the line of sight, an effect
that is again missed by our single-density assumption.
As mentioned above, fixing these fitting imperfections would require
having a realistic description of the multiple density regimes present along 
any line of sight. Lacking such a description, we consider that our single-density 
model provides a reasonably good fit to the observed intensities.

As a final illustration of our modeling, we present in
Fig.~\ref{lvg_n2hp_c2h} the solutions
for N$_2$H$^+$ and C$_2$H, the
two species that, together with CN, require 
special abundance profiles.
The top panel of the figure compares the N$_2$H$^+$ data with the prediction from two abundance
models. The dashed red line corresponds to a constant-abundance model set to fit
the observed intensities at high $N$(H$_2$).
As can be seen, the model significantly overestimates the intensities in the outer cloud,
indicating the need of an abundance drop at low $N$(H$_2$).
The second model (dashed blue line) corresponds to the
profile described in Appendix~\ref{app_abu}, and has 
a significant abundance drop at low H$_2$ column densities,
as expected from the destruction of N$_2$H$^+$ by CO when the latter species
is abundant in the gas phase \citep{ber97,aik01}.
This modified model reproduces the emission both at high and low H$_2$
column densities, and is therefore in better agreement with the observations. 
It should be noted, however, that our data sampling 
in the transitional  region ($N(\mathrm H_2) \approx 10^{22}$~cm$^{-2}$)
is not fine enough to constrain well the abundance drop, and that the
drop could be sharper than suggested by the model. 
Additional observations of the N$_2$H$^+$ emission 
in this transitional region and at lower column densities are needed to
fully characterize the distribution of this unique dense gas tracer
through the entire cloud.

The bottom panel of Fig.~\ref{lvg_n2hp_c2h} shows the model results
for the C$_2$H(1--0) line. Again, the panel compares two abundance models 
with the survey data.
The dashed red line represents a standard
abundance model that has both outer photodissociation and inner freeze-out 
contributions.
This model fits the emission in the inner cloud, but fails to reproduce
the observations at very low column densities.
The dashed blue line
corresponds to an abundance profile that has an outer
enhancement next to the photodissociation edge, and that is
inspired by the PDR model of the C$_2$H abundance from \cite{cua15} 
(see Appendix~\ref{app_abu} for a full description). 
As can be seen, this modified model reproduces
better the emission enhancement near the outer edge of the cloud in
addition to the cloud interior.
A similar model, but with a smaller outer enhancement, is also
needed to fit the CN emission, as shown in Appendix~\ref{sect_remaining}.

We summarize our modeling results by saying that
the Perseus data suggest that the shape of the abundance profile of 
any species is mostly controlled by how the species
reacts to two agents: (i) the UV ISRF at low column densities 
and (ii) dust collisions at high column densities.
All our survey species except N$_2$H$^+$, C$_2$H, and CN behave passively
with respect to these agents, in the sense that they are photodissociated by
the UV radiation and they freeze out onto the dust grains.
As a result, the abundance of these species decreases both toward the outer edge and 
inside of the cloud in a manner illustrated by the top panel of Fig.~\ref{abundances}.

The three exceptions we found to the above abundance pattern result from some type 
of active reaction to either the UV radiation or to the collisions with the dust.
In the case of C$_2$H and CN, these species 
are enhanced near the cloud edge as a result of the UV ISRF, and in the case of
N$_2$H$^+$, this molecule thrives when CO starts to freeze out.
These two behaviors are illustrated in the middle and bottom panels
of Fig.~\ref{abundances}. When taken into account,
the complete set of observed lines in Perseus 
can be reproduced with a relatively simple model.

The fact that a simple model like the one presented here 
can reproduce the line intensities of so many species and transitions
suggests that the main properties of the line emission
in the Perseus cloud are controlled by a small number of processes  
that can be simulated with a few model parameters.
Whether this behavior is peculiar to Perseus or common to other
clouds requires further investigation, and will be explored in future work.

\section{Discussion}

\subsection{Origin of the intensity dependence with column density}
\label{origin}

Since our cloud model reproduces the main features of 
the Perseus emission, we can use it to investigate
the origin of the different correlations between 
intensity and H$_2$ column density found for different species. 
Of particular interest is the comparison between
the correlation of the CO isotopologs, which present a rather flat
dependence with $N$(H$_2$), and the correlation of the traditional 
dense gas tracers
HCN and CS, which approximately follow a linear 
dependence with $N$(H$_2$).

\begin{figure}
 	\centering
 	\includegraphics[width=0.8 \hsize]{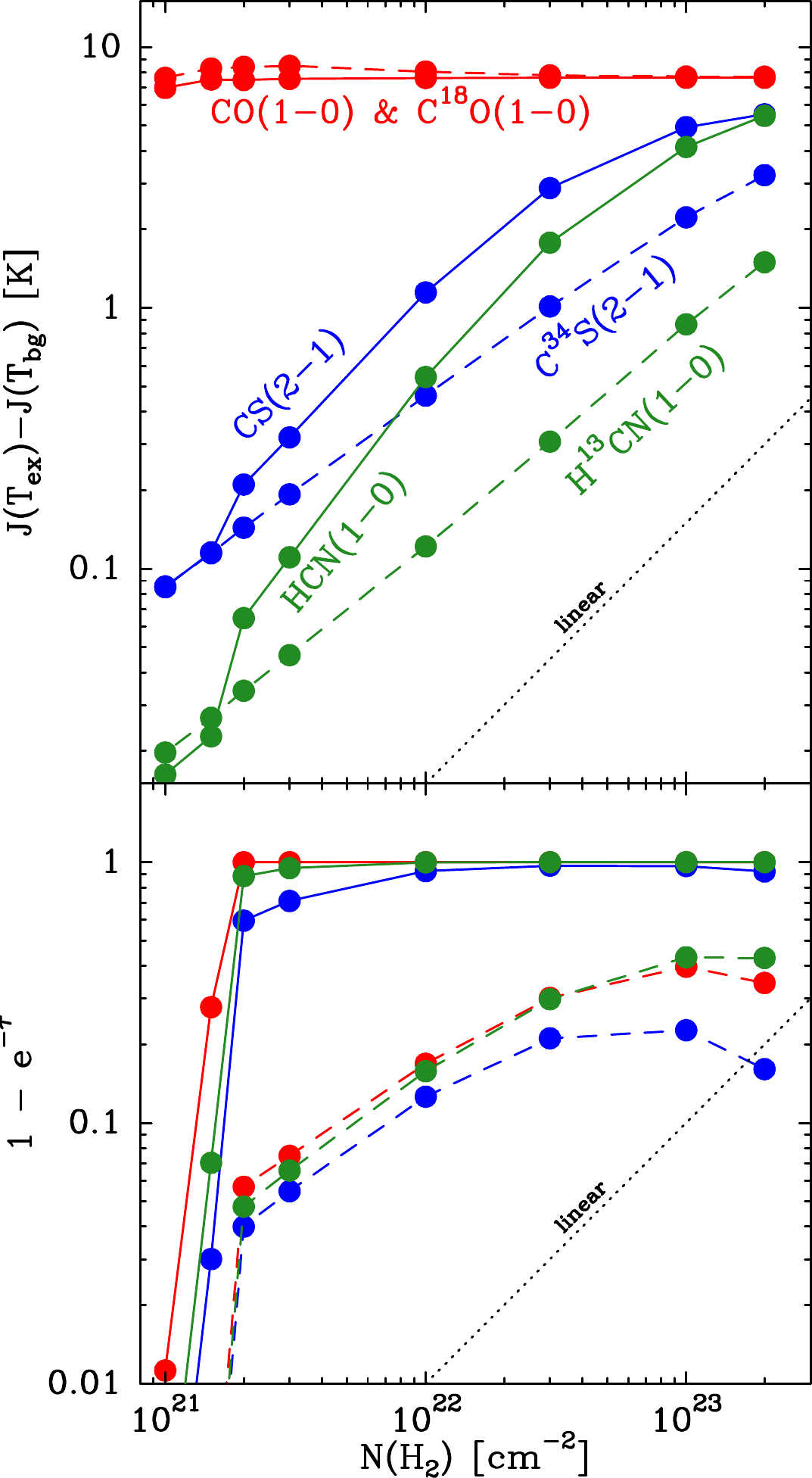}
 	\caption{Model results for the two factors of the radiative transfer solution:
 		excitation term {\em (top)} and optical depth term {\em (bottom)}.
 		The solid lines represent the results for the main isotopologs
 		of CO (red), HCN (green), and CS (blue). The dashed lines represent 
 		equivalent results for the
 		rare species (C$^{18}$O, H$^{13}$CN, and C$^{34}$S).}
 	\label{tex_tau}
 \end{figure}

To investigate the origin of these different correlations,
we look separately at the two factors that contribute to the
intensity in the solution of the equation of radiative transfer:
$J(T_{\mathrm{ex}}) - J(T_{\mathrm{bg}})$ and  $(1-e^{-\tau})$,
which we will refer to as the ``excitation'' factor and the 
``optical depth'' factor, respectively.
In Fig.~\ref{tex_tau} we present their value as a function of $N$(H$_2$) 
for the 3mm lines of the main CO, HCN, and CS isotopologs (solid lines) 
and the rare isotopologs C$^{18}$O, H$^{13}$CN, and C$^{34}$S (dashed lines).

As Fig.~\ref{tex_tau} shows, the excitation factor for both the 
main and rare CO isotopologs 
(top panel, solid and dashed red lines) has an approximately constant 
value close to the 
gas kinetic temperature minus the background temperature, 
as expected for species that are thermalized over most of the cloud.
The optical depth factor, on the other hand,
is different for the main and rare CO isotopologs
(bottom panel, red lines).
For the main isotopolog, the optical depth
factor has a close-to-constant value of 1 over most of the cloud due to the 
extremely high optical depth of the emission. 
For the less abundant C$^{18}$O, the optical depth factor
reflects closely the C$^{18}$O column density, which
is not linear with $N$(H$_2$), but curves downward at high values
due to the increasing effect of freeze out.
When the excitation and optical depth factors
are multiplied to obtain the intensity, 
the result is an approximately constant function for CO and 
a slightly curved intensity law for C$^{18}$O, as
observed in the data.

For the HCN and CS isotopologs, the excitation factor
differs sharply from that of CO. As can be seen 
from the green and blue lines in the top panel of Fig.~\ref{tex_tau},
while the excitation factor of CO is flat, that of HCN and CS
increases rapidly with $N$(H$_2$).
This increase is a consequence of the HCN and CS lines being significantly subthermal,
and therefore having an excitation that increases rapidly as the volume density increases
with $N$(H$_2$).
For the main HCN and CS isotopologs (solid lines), the excitation has
an additional contribution from photon 
trapping due to the high optical depth of the lines,
and this enhances their factor over that of the rare isotopologs,
for which trapping is insignificant (dashed lines).
Independently of this extra contribution, the
excitation factor of all the HCN and CS isotopologs
increases by about two orders of magnitude over the H$_2$ column density
range of the cloud. 

In contrast with the excitation factor, the
optical depth factor of HCN and CS
behaves like that of CO (bottom panel).
For the main isotopologs, the factor has an almost
constant value of one inside the photodissociation edge,
while for the less abundant isotopologs, it
presents a less-than-linear increase with $N$(H$_2$)
due to the effect of freeze out.
This similarity with CO is not surprising
since all the species have similar abundance profiles,
and the main isotopologs are very optically thick while the
rare ones are thin.

The similar behavior of the optical depth factors
of CO and the traditional dense gas tracers indicates
that the different dependence of the intensity  
of HCN-CS and CO with $N$(H$_2$) 
in the cloud interior
arises mainly from a difference in the
excitation of the molecules. For the CO isotopologs, the combination of 
flat excitation factors due to thermalization 
with flat or not too steep optical depth
factors results in relatively flat intensity correlations with
$N$(H$_2$). For HCN and CS, the steep excitation factors
resulting from subthermal excitation
dominate the dependence of the emergent intensities 
and are ultimately responsible for the strong correlation of
the intensities with $N$(H$_2$). 

To understand why the intensity of the
main and rare isotopologs of HCN and CS
present similar dependence with $N$(H$_2$), we need to consider now
the combined effects of excitation and optical depth.
As seen in the top panel of Fig.~\ref{tex_tau}, the main isotopologs 
(solid lines) have significantly steeper excitation factors than 
the rare isotopologs (dashed lines) 
due to the additional contribution from photon trapping.
These factors have slopes that, although not constant,
are close to linear, as can be seen from a comparison with the
dotted line shown in the panels.
When these factors are multiplied by the almost constant  optical 
depth factors, the resulting intensities retain the close-to-linear slope.

The rare isotopologs, on the other hand, present slightly 
flatter excitation factors
due to the missing trapping contribution. When these factors are
multiplied by the optical depth factors, which have a non-negligible 
slope over most of the $N$(H$_2$) range, the resulting emergent 
intensity approaches the linear slope of the thick main isotopologs.
It seems therefore that the similar behavior
of the thin and thick traditional dense gas tracers
as a function of H$_2$ column density
results from the  approximate compensation of their
different excitation and optical depth factors.
While somewhat fortuitous, this behavior seems to be very robust
since it is displayed by most observed species
in multiple transitions, and ultimately gives rise to the 
systematic quasi-linear emission pattern found by 
most dense gas tracers in our survey.

\subsection{What the traditional dense gas tracers trace}
\label{sect_tracers}

The almost linear dependence with H$_2$ column density
of the emission from species like HCN and CS  forces us to reevaluate the
traditional distinction between dense gas tracers and column density tracers
used to classify molecules.
The correlations found in Perseus imply that,
strictly speaking, only N$_2$H$^+$ deserves the term dense gas tracer
since it is the only species truly selective of the dense gas:
its emission remains undetected below around $10^{22}$~cm$^{-2}$
(corresponding in our model to a density of $2\times 10^4$~cm$^{-3}$),
and then rises nonlinearly when $N$(H$_2$) exceeds the threshold.
This unique behavior of N$_2$H$^+$ has been previously seen at core scales 
\citep{cas99,ber02,taf02} and at cloud scales \citep{kau17,pet17}, and 
results from the sensitivity of N$_2$H$^+$
to the presence of CO in the gas phase.
Although no other molecule in our survey presents a similar behavior,
it is likely that NH$_3$ may do so since
NH$_3$ shares with N$_2$H$^+$ a resilience to
freeze out in addition to being a late-time chemical
species \citep{suz92,taf02}. Indeed, 
observations of N$_2$H$^+$ and NH$_3$ toward Perseus dense
cores have found a strong correlation between
the emission of these two species \citep{joh10,hac17}.
Extending our Perseus observations to include the cm-wavelength 
NH$_3$ lines would be highly desirable to test this prediction.

Although HCN, CS, and the other traditional dense gas tracers
do not display the selectivity to the dense gas of N$_2$H$^+$,
they are still sensitive to the cloud dense material
in an indirect way. 
This is so because their emission follows the column density,
which we have seen is itself an 
indirect indicator of the gas volume density.
These tracers, therefore, are brighter toward the higher column density
gas, which results from the presence of dense gas along the line of sight. 
For this reason, tracers like HCN and CS are  still
useful to identify the densest parts
of molecular clouds in spatially resolved maps, as
proven by numerous previous studies (e.g., \citealt{plu97, beu02, wu10}).

The lack of selectivity to the dense gas of HCN, CS, and similar tracers
is a more serious problem in
unresolved observations of molecular clouds, as those commonly used
in extragalactic work.
In this type of work, interpreting the emission from the different tracers 
is not straightforward
since this emission represents a weighted average to which all
the density regimes in the cloud contribute.
The relevant parameter in this case is
the product of the column density probability
distribution function (PDF) times the line intensity as a function of column 
density. As we have seen, the Perseus data show that for HCN, CS, and other tracers,
the intensity
is approximately linear with $N$(H$_2$), while the column density
PDF depends much more steeply on the parameter, as $N$(H$_2$)$^{-3}$ \citep{zar16}.
As a result, the integrated emission will be 
dominated by the contribution from the lowest column density positions, whose
nonlinear overabundance more than compensates for the approximately
linear decrease in their intensity.

A similar dominant contribution of the low-density gas in the cloud-integrated 
emission from the traditional dense gas tracers
has also been found by several studies, and likely 
represents a general trend among molecular clouds 
\citep{kau17, pet17, wat17, shi17,nis17,eva20}.
This finding should be not surprising if in most clouds, as in Perseus,
the PDF has a very steep negative dependence with $N$(H$_2$)
\citep{lom15}, while the line intensities depend almost linearly with $N$(H$_2$)
(e.g., \citealt{pet17}).
These two behaviors naturally combine to make the contribution of the
low column density positions increasingly more important in the
cloud-integrated intensity.

Even if the integrated emission of the traditional dense gas tracers
is dominated by the low-density gas,
it still has a relevant physical interpretation.
We have seen that in Perseus the emission follows
quasi-linearly the column density in the region
above the photodissociation threshold. As a result, 
the cloud-wide integrated intensity, which is obtained integrating 
the product of the PDR times the intensity as a function of the column density,
will be proportional to the mass of the cloud above the photodissociation threshold.
Since our models indicate a threshold of about 
$A_{\mathrm V} = 2$~mag for most species, 
the cloud-integrated intensity of these tracers will
be proportional to the mass of the gas
above two magnitudes of $A_{\mathrm V}$.

While no other cloud has been characterized using 
the sampling method we used for
Perseus, we have seen that an approximately linear dependence
of the intensity with the column density is a likely
feature of several other clouds (Sect.~\ref{sect_compare}).
If this is so in general, the same relation between
cloud-integrated intensity and mass above the
photodissociation threshold of a cloud will apply.
Such a possibility, which clearly requires further investigation,
is of interest for extragalactic work, 
where there is evidence for a linear correlation
between the emission of tracers such as HCN(1--0) 
and the rate of star formation \citep{gao04, ken12, use15, jim19}.
Interpreting the observed line intensities as estimates of
the cloud mass inside a certain column density threshold could 
potentially help explain the origin of the correlation.

Given the above, it is worth noting that
\cite{lad10} have found that
for local clouds (including Perseus),
the mass above a column density threshold
of $A_{\mathrm V} \approx 7.3$~mag ($A_{\mathrm K} \approx 0.8$~mag) 
is linearly proportional to
the star-formation rate of the cloud.
This 7.3~mag threshold is significantly higher than the 
photodissociation threshold we found in Perseus, so the mass derived using 
the \cite{lad10} threshold
is different from (and much smaller than) the mass inferred from 
the HCN(1--0) intensity\footnote{The 7.3~mag mass will be close to 
that inferred from the N$_2$H$^+$ intensity,
which becomes observable at column densities  
equivalent to about $A_{\mathrm V} = 10$~mag, and is
in fact a favored tracer of star-forming sites in local clouds.}.
It is therefore unclear whether the \cite{lad10} relation
can help connect the cloud-integrated HCN(1--0) intensity with
the star-formation rate as found
by extragalactic observations. One possibility is that the
masses set by the 2 and 7.3~mag thresholds are linearly correlated, 
something that is possible given the quasi-universal power-law
nature of the column density PDFs \cite{lom15}.

The above speculations clearly show that
an observational effort is needed to characterize the emission from multiple 
clouds and test whether our Perseus results reflect a more general trend. 
For such an effort, the stratified random sampling method presented here
appears to be an ideal tool.

\section{Conclusions}

We used stratified random sampling to select 
100 target positions that represent the different 
regimes of the H$_2$ column density in the Perseus molecular cloud. 
We observed these positions with the IRAM 30m telescope covering the 
3mm-wavelength band and selected parts of the 2mm and 1mm bands.
We studied the correlation of the observed line intensities 
with the H$_2$ column density, and we developed a simple cloud model 
to reproduce the main features of the emission. 
The main conclusions from this work are the following.

1. A comparison of our sampling results for $^{12}$CO(1--0) and $^{13}$CO(1--0) with the
mapping data from the COMPLETE project shows that stratified random sampling
can be used to estimate the mean intensity and the intensity dispersion 
of these two lines as a function of column density within a factor of 1.5 or
better. 

2. The intensity of the molecular lines in Perseus  
strongly correlates with the H$_2$ column density 
over the two orders of magnitude that this quantity spans through the cloud.
The lines of the CO isotopologs and the traditional
dense gas tracers (CS, HCN, HCO$^+$,
etc.) present a level of dispersion in the intensity-$N$(H$_2$) relation of only 
about 0.2 dex. This tight correlation between line intensities and H$_2$ column density 
supports the use of column density as a proxy  of the intensity for the stratified random 
sampling, 
as initially motivated by principal component analysis \citep{ung97,gra17}.

3. The intensity of the CO isotopologs increases slower than linearly with $N$(H$_2$),
while the intensity of most dense gas tracers increases approximately linear with 
$N$(H$_2$).

4. Several dense gas tracers present significant deviations from a linear dependence 
with $N$(H$_2$).
The largest deviation is that of N$_2$H$^+$(1--0), which presents
a very rapid transition from undetected to relatively bright near 
$N$(H$_2$) = $10^{22}$~cm$^{-2}$. 
C$_2$H(1--0) and (to smaller extent) CN(1--0) present significant enhancements 
in their intensity at low column densities (< $10^{22}$~cm$^{-2}$).

5. The main emission trends identified in the Perseus data
are similar to those recently reported from 
multiline mapping of entire clouds (e.g., \citealt{kau17, pet17, wat17}).
This similarity suggests that the sampling technique can truly characterize 
efficiently
the emission from a cloud. It also points to a general
behavior of the emission in clouds that span
a relatively large range of star-formation conditions.

6. A simple cloud model can reproduce the main emission features of Perseus. 
For most species, the model requires a molecular abundance that has
a sharp drop at low column densities (likely due to photodissociation)
and a more gradual decrease at high column densities (likely caused by freeze out).
The only species that require different abundance distributions are N$_2$H$^+$, 
which is enhanced when CO freezes out, and C$_2$H and CN, which are enhanced by the
external UV field. The different abundance behaviors of these species 
are consistent with our current understanding of molecular cloud chemistry.

7. Our cloud model suggests that the flat dependence with $N$(H$_2$) of the 
CO isotopologs
results from a combination of optical depth effects and molecular freeze out under
thermalized conditions. The quasi-linear behavior of most traditional
dense gas tracers and their isotopologs results from the increase in excitation
with column density combined with molecular freeze out toward the densest gas.
A lucky compensation between excitation in the optically thick species and
column density effects in the rare isotopologs is responsible for their
similar behavior with $N$(H$_2$).

8. The cloud-integrated emission of the traditional dense gas tracers
is dominated by the contribution from 
the lowest column density positions because their nonlinear overabundance (described by the
PDF) more than compensates for the approximately linear decrease in their intensity.
The quasi-linear dependence of their intensity with $N$(H$_2$) makes the integrated intensity
of these tracers approximately 
proportional to the mass of the gas interior to the photodissociation edge.
This property may help provide a physical meaning to the integrated intensity of
these tracers in unresolved observations.

\bigskip

\begin{acknowledgements}
 We thank an anonymous referee for a thorough review of the manuscript and
for numerous comments and suggestions that have helped improve this work.
We thank the IRAM staff for their excellent support during the 30m telescope
observations.
We acknowledge support from grant AYA2016-79006-P from MINECO, which
is partly funded through FEDER, and from grant PID2019-108765GB-I00.
AU acknowledges support from  PGC2018-094671-B-I00 (MCIU/AEI/FEDER).
AH acknowledges support from VENI project
639.041.644, which is (partly) financed by the Netherlands Organisation for
Scientific Research (NWO).
This work is based on IRAM 30m-telescope observations carried out under project
numbers 034-17 and 104-17.
IRAM is supported by INSU/CNRS (France), MPG (Germany), and IGN (Spain).
This research has made use of NASA’s Astrophysics Data System Bibliographic Services and the SIMBAD database, operated at CDS, Strasbourg, France.

\end{acknowledgements}

\bibliographystyle{aa} 
\bibliography{pers_sampl.bib}

\appendix

\section{Sample positions and 3mm line intensities}
\label{app_positions}

Table~\ref{tbl_master} presents the coordinates, column density estimates, and 3mm line
intensities for the 100 sample positions used in the Perseus survey. A full version of 
the table is available on-line.

\begin{sidewaystable*}
	\caption[]{Sample positions and line intensities.\label{tbl_master}}
	\centering
	\begin{tabular}{lccccccccccc}
		\hline\hline
		\noalign{\smallskip}
		Position\tablefootmark{a} & RA(J2000) & Dec(J2000) & $N$(H$_2$) & $^{12}$CO(1--0) & $^{13}$CO(1--0) & C$^{18}$O(1--0) & C$^{17}$O(1--0) & HCN(1--0) &  CS(2--1) & HNC(1--0) & HCO$^+$(1--0) \\
		
		& ($^\mathrm{h}$~~$^\mathrm{m}$~~$^\mathrm{s}$) & (\degr~~\arcmin~~\arcsec) & (cm$^{-2}$) & (K km s$^{-1}$) & (K km s$^{-1}$) & (K km s$^{-1}$) &  (K km s$^{-1}$) & (K km s$^{-1}$) &
		(K km s$^{-1}$) & (K km s$^{-1}$) & (K km s$^{-1}$) \\
		\hline
		\noalign{\smallskip}
		
		PERS-10\_01 & 03~29~04.7 & +31~14~44.8 & $1.0\; (0.1) \times 10^{23}$ &  118 (12) &  30 (3) &  4.6 (0.5) &  1.3 (0.1) & 12 (1) & 9.4 (0.9) & 9.5 (1.0) & 10.0 (1.0)   \\
		PERS-10\_02 & 03~29~09.4 & +31~13~35.6 & $1.6\; (0.3) \times 10^{23}$ & 82 (8) & 25 (3) & 3.6 (0.4) & 1.1 (0.1) & 13 (1) &  11.4 (1.1) & 9.2 (0.9) & 7.8 (0.8) \\
		PERS-10\_03 & 03~29~16.0 & +31~12~29.6 & $1.1\; (0.1) \times 10^{23}$ & 73 (7) & 22 (2) & 3.2 (0.3) & 0.88 (0.09) &4.9 (0.5) & 4.4 (0.4) & 4.1 (0.4) & 3.4 (0.3) \\
		PERS-10\_04 & 03~33~16.7 & +31~07~18.0 & $1.2\; (0.1) \times 10^{23}$ & 51 (5) & 21 (2) & 5.6 (0.6) & 2.0 (0.2) & 4.7 (0.5) &  4.9 (0.5) &  4.7 (0.5) & 2.6 (0.3) \\
		PERS-10\_05 & 03~25~36.0 & +30~45~57.2 & $1.2\; (0.2) \times 10^{23}$ & 49 (5) & 19 (2) & 3.2 (0.3) &  0.92 (0.1) & 11 (1) & 6.2 (0.6) & 9.7 (1.0) &  7.8 (0.8) \\
		PERS-10\_06 & 03~33~15.8 & +31~07~03.9 & $1.3\; (0.1) \times 10^{23}$ & 48 (5) & 20 (2) &  5.5 (0.6) &  2.2 (0.2) & 4.0 (.4) & 3.8 (0.4) &  3.7 (0.4) & 1.7 (0.2) \\
		PERS-10\_07 & 03~29~02.2 & +31~15~32.1 & $1.1\; (0.1) \times 10^{23}$ & 105 (10) & 36 (4) & 5.2 (0.5) & 1.5 (0.2) & 11 (1) & 9.7 (1.0) &  9.3 (0.9) & 14.2 (1.4) \\
		PERS-10\_08 & 03~29~10.8 & +31~14~29.3 & $1.1\; (0.1) \times 10^{23}$ &  73 (7) &  23 (2) & 3.2 (0.3) & 0.75 (0.08) & 8.1 (0.8) & 7.3 (0.7) & 6.8 (0.7) &  5.7 (0.6) \\
		PERS-10\_09 & 03~29~09.7 & +31~14~40.3 & $1.3\; (0.1)\times 10^{23}$ & 78 (8) & 23 (2) &  2.9 (0.3) &  0.86 (0.09) & 9.0 (0.9) & 7.1 (0.7) & 8.1 (0.8) & 6.2 (0.6) \\
		PERS-10\_10 & 03~29~13.5 & +31~13~16.8 & $1.5\; (0.3) \times 10^{23}$ & 72 (7) &   24 (2) &  3.7 (0.4) &  1.2 (0.1) & 5.5 (0.6) & 5.4 (0.5) & 4.5 (0.4) &  3.9 (0.4) \\		
		\hline
		\noalign{\smallskip}
	\end{tabular}
	\tablefoot{A full version of this table is available on-line.
		\tablefoottext{a}{The first number in the position name indicates the column density bin and the second one indicates the order in our sampling sequence.}
	}	
\end{sidewaystable*}

\section{Stacked spectra}
\label{app_stck}

\begin{figure*}
	\centering
	\includegraphics[height=23cm]{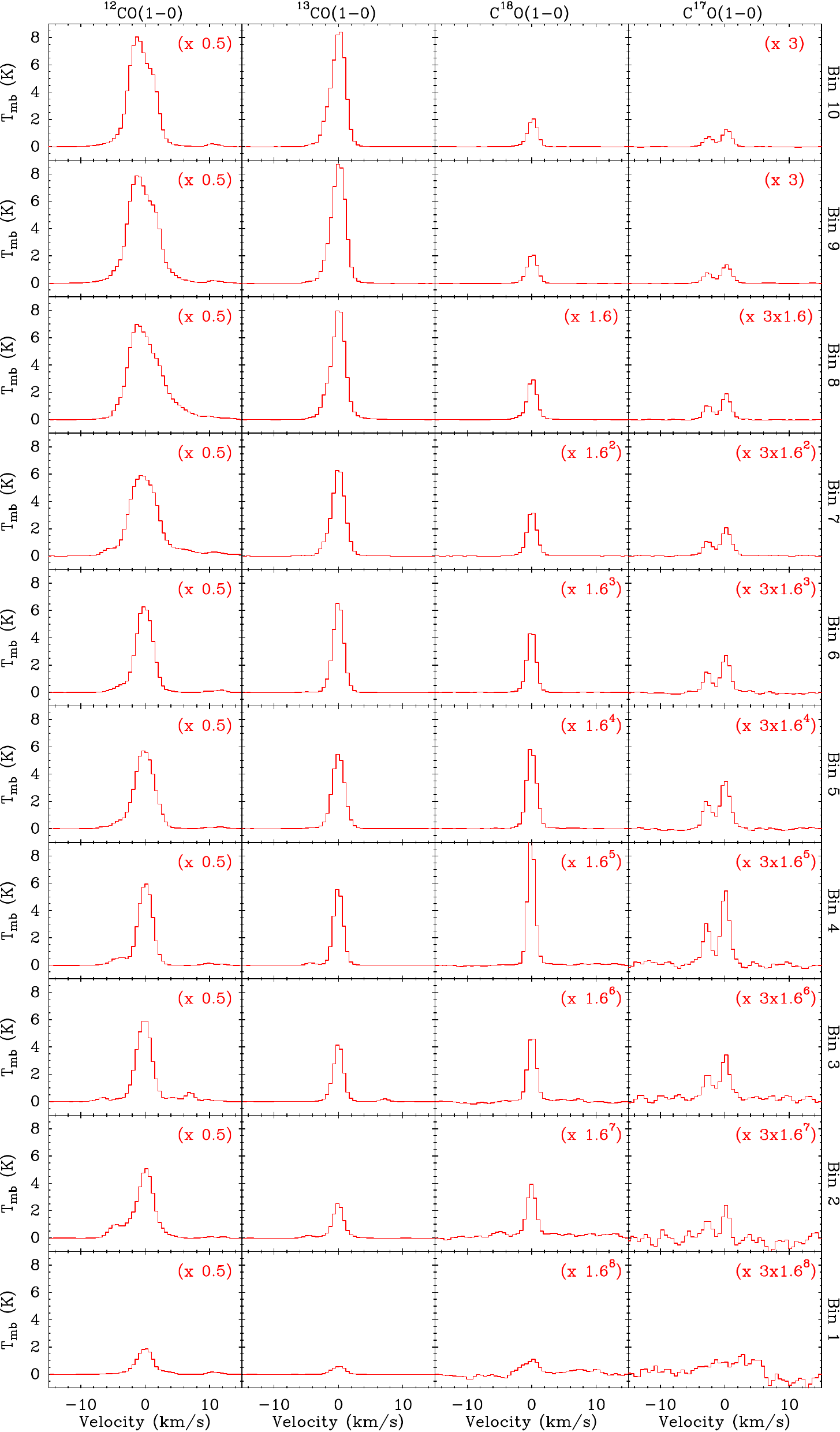}
	\caption{Stacked spectra of the $J$=1--0 transition for different CO isotopologs 
		Each spectrum represents the average of the ten spectra taken in the column-density
		bin indicated in the right label. The mean column density of these bins decreases downward
		by factors of 1.6, and the spectra have been scaled up by different factors to
		approximately maintain the same physical size.}
	\label{fig_stck_co}
\end{figure*}

\begin{figure*}
	\centering
	\includegraphics[height=23cm]{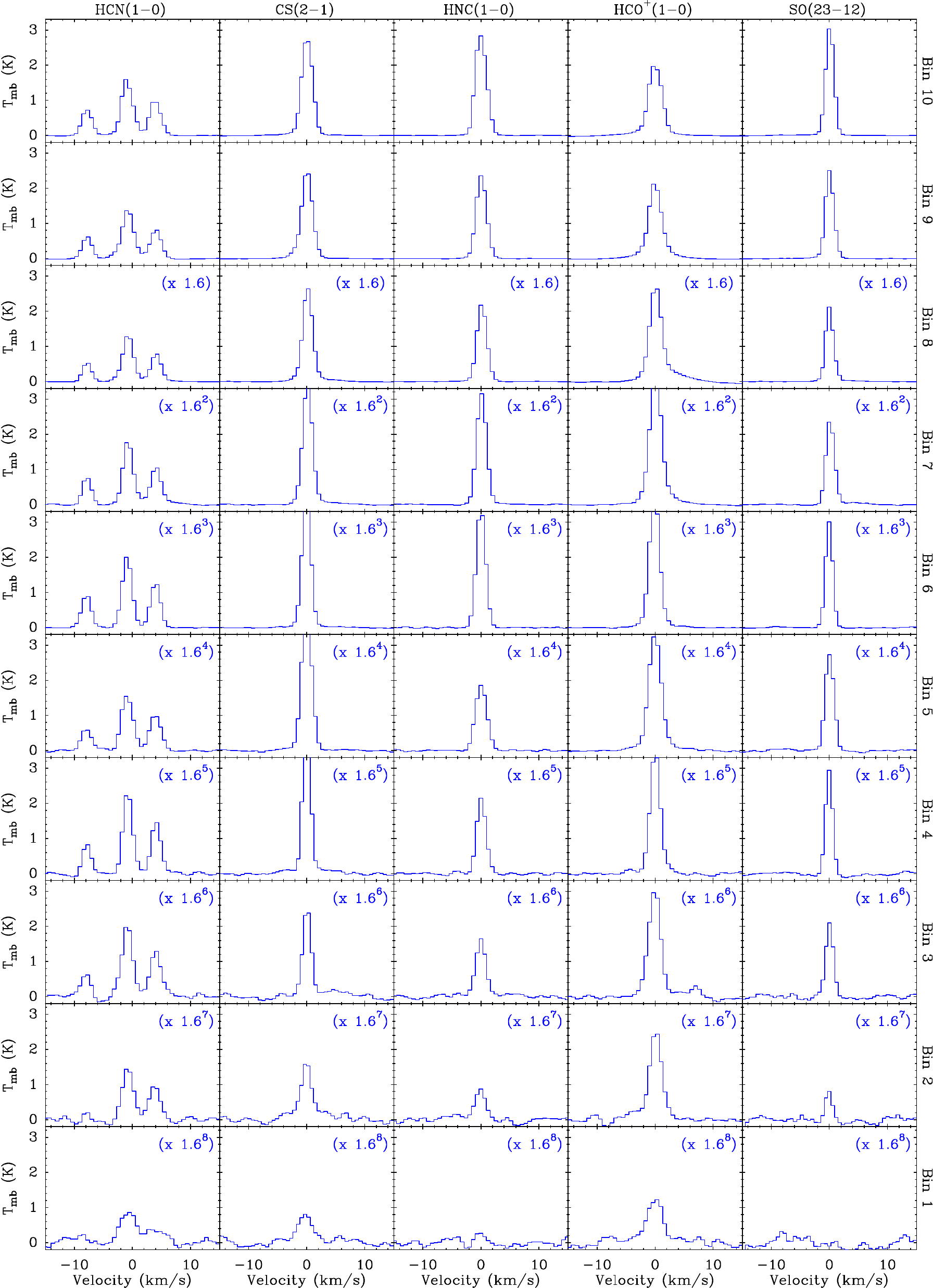}
	\caption{Same as Fig.~\ref{fig_stck_co} but for the traditional dense gas tracers: HCN(1--0), CS(2--1), HNC(1--0), HCO$^+$(1--0), and SO(23--12).}
	\label{fig_stck_dense}
\end{figure*}

\begin{figure*}
	\centering
	\includegraphics[height=23cm]{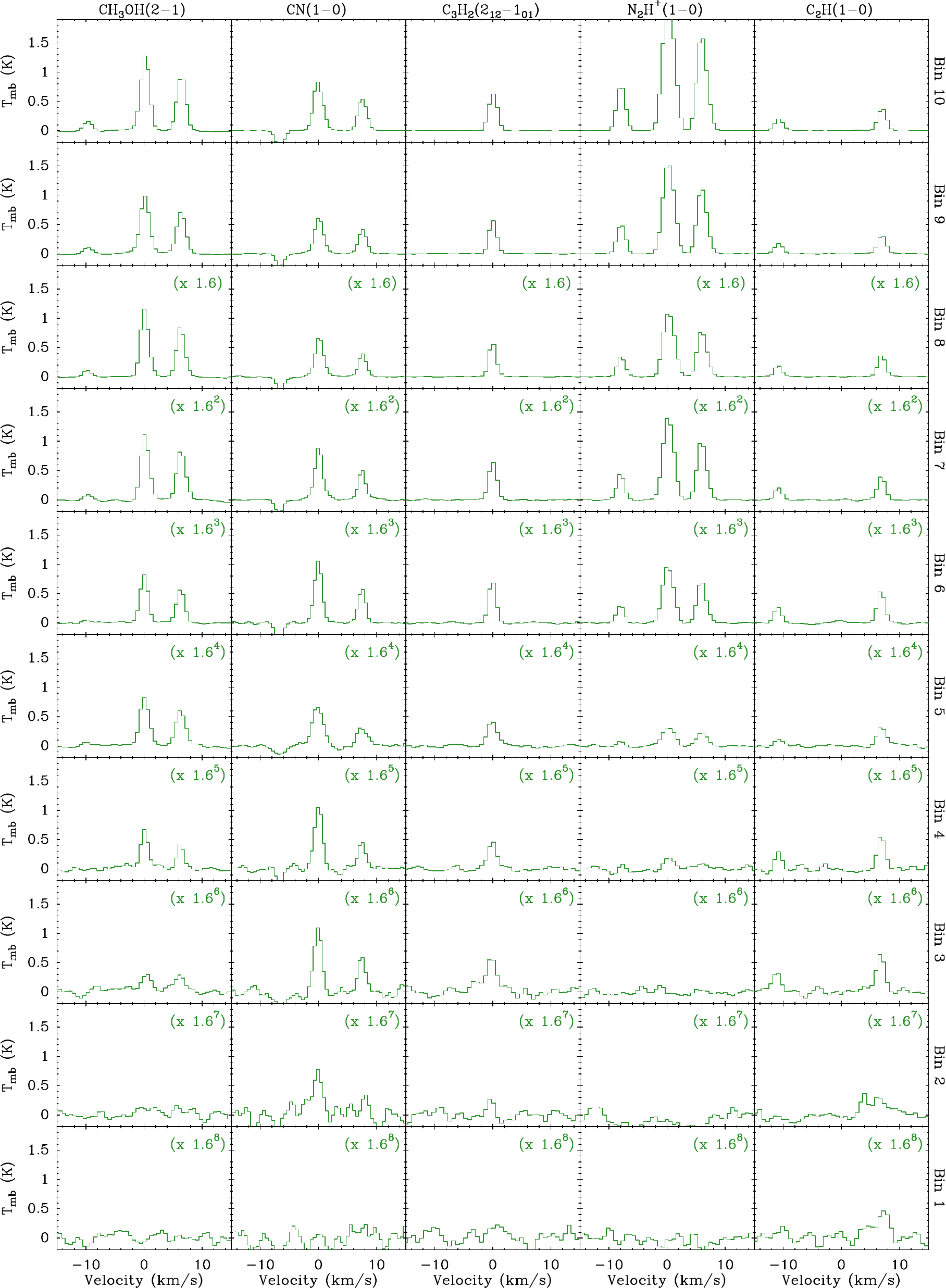}
	\caption{Same as Fig.~\ref{fig_stck_co} but for the additional gas tracers: CH$_3$OH(2--1), CN(1--0), C$_3$H$_2$(2$_{12}$--1$_{01}$), N$_2$H$^+$(1--0), and C$_2$H(1--0). For CN(1--0) and C$_2$H(1--0),
	only a subset of the hyperfine components is shown due to the limited velocity window.
	The negative CN(1--0) feature near -7~km~s$^{-1}$ is an artifact from the frequency 
	switching observing mode.}
	\label{fig_stck_others}
\end{figure*}

 Figs.~\ref{fig_stck_co}, \ref{fig_stck_dense}, and \ref{fig_stck_others} present 
stacked spectra for the 3 mm-wavelength transitions discussed in Sect.~\ref{sect_over}.
The stacking was done by averaging the ten spectra of each
column density bin after having shifted the lines to zero velocity using a Gaussian fit to the $^{13}$CO(1--0) line. To ease inter-comparing
the spectra, each plot uses the same intensity scale 
(in units of $T_{\mathrm{mb}}$). Most spectra 
have been scaled by factors of $1.6^n$, where $n$ depends on the bin number,
to approximately compensate for
the change in H$_2$ column density between the bins.

\section{Modeling abundance profiles}
\label{app_abu}

As mentioned in Sect.~\ref{sect_abu}, our cloud model aims to provide a description of the
abundance profiles that is both simple and consistent with 
the chemical processes known to occur in a cloud. 
To this end, we found convenient to write
the abundance profile of each species (with respect to H$_2$) 
as the product of three factors,
$$X({N(\mathrm{H_2})}) = X_0 \times f_{\mathrm{out}}(N({\mathrm{H_2}})) 
\times f_{\mathrm{in}}(N({\mathrm{H_2}})),$$
where $X_0$ is a constant scaling, $f_{\mathrm{out}}$ represents
a departure from constant  near the outer edge of the cloud, likely caused
by the UV photons from the interstellar radiation field,
and $f_{\mathrm{in}}$ represents a departure from constant 
at high column densities, likely related to molecular freeze out.

\begin{figure}
   \centering
   \includegraphics[width=\hsize]{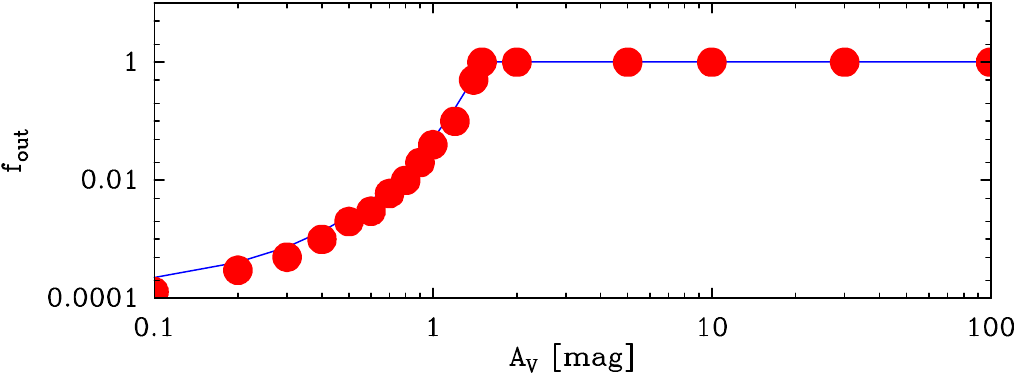}
   \caption{Abundance factor used to model the cloud photodissociation edge. 
   	The red symbols represent values of the CO abundance predicted by the F1 
   	PDR model of \cite{rol07} and extracted from their Fig.~4b.
   	The solid blue line is the analytic expression described in the text
   	for a choice of $A_0 = 1.5$~mag.}
   \label{f_pdr}
\end{figure}

For all species but C$_2$H and CN, the $f_{\mathrm{out}}$ factor 
needed to fit the data represents 
a sharp abundance drop that we associate with
the photodissociation of the molecules 
by the external UV radiation field. Although the detailed shape of this
drop is not very critical for our modeling due to the lack of
column-density 
resolution near the cloud edge, we used a realistic
profile based on the photon dominated region (PDR) models of \cite{rol07}.
As a representative profile, we selected the
CO abundance curve estimated by these authors in their F1
model ($n$(H$_2$) = $5 \times 10^3$ cm$^{-3}$, and 
$\chi = 10$). 
This model has a Draine $\chi$ parameter \citep{dra78} that is intermediate
between the values favored by \citet{pin08} from their fit of the Perseus CO lines
($\chi=1$-3) and the recent determination for B1 by \citet{nav19} using a
a model of the dust temperature ($\chi=24$).
We fitted this profile
using the following analytic expression
$$
f_{\mathrm{out}}(N({\mathrm{H_2}})) =
\begin{cases}
%[\exp(A_V-A_0)]^6 & {\mathrm{if}}~~~A_V \le A_0 \\
[e^{(A_\mathrm{V}-A_0)}]^6 & {\mathrm{if}}~~~A_\mathrm{V} \le A_0 \\
1 & {\mathrm{if}}~~~A_\mathrm{V} > A_0, 
\end{cases}
$$
where $A_\mathrm{V}$ is the H$_2$ column density expressed in units of
visual extinction ($N$(H$_2$) = $9.35 \times 10^{20} A_\mathrm{V}$ cm$^{-2}$,
\citealt{boh78}).
$A_0$ is a free parameter that describes
the location of the photodissociation edge also in units of
visual extinction and allows for changes in the value of $\chi$
with respect to the original model.
Fig.~\ref{f_pdr} compares a set of  abundance values extracted from 
Fig.~4b in \cite{rol07} (red circles) with
our analytic expression
for a choice of $A_0 = 1.5$~mag  (blue line).
As shown below in Table~\ref{tbl_abu}, the $A_0$ values
needed to fit the Perseus data range between 1 and 2~mag,
in good agreement with the theoretical expectation from the
PDR model.
It should be noted, however, that for most species, our observations
	 cannot constrain the exact location
	of the photodissociation edge due to the limited signal to noise ratio of the intensities
	at low column densities and the coarseness of the column density bins. 
	For these species, we used
	a single value of 2~mag based on the results from the CO data. This choice
provides a reasonable match to the data, but it should not 
be considered as a well-defined best fit.

Since the C$_2$H and CN intensities require a significant abundance enhancement in the
vicinity of the photodissociation edge, 
we multiplied the $f_{\mathrm{out}}$ factor of these species
by the additional term

$$f'_{\mathrm{out}}(N({\mathrm{H_2}})) = 1 + \alpha\; e^{-A_\mathrm{V}/3},$$
where the free parameter $\alpha$ describes the magnitude of the 
outer
abundance enhancement. This additional factor was inspired by the
detailed modeling of the C$_2$H abundance in the Orion Bar PDR
by \cite{cua15}, who found that the species undergoes 
an abundance enhancement of several orders of magnitude
in the vicinity of the cloud edge (their Fig. 17).

The last factor in our abundance parameterization is $f_{\mathrm{in}}$,
which describes deviations from a constant value in the cloud interior. 
For all species but N$_2$H$^+$, the data requires a significant
abundance drop at high column densities that is likely caused by freeze out onto
the dust grains.
Following previous freeze-out modeling by \cite{taf02}, we 
described this abundance drop with the simple analytic expression
$$f_{\mathrm{in}}(N({\mathrm{H_2}})) = e^{-n(\mathrm{H}_2)/n_{\mathrm{fr}}},$$
where $n(\mathrm{H_2})$ is the gas density (related to $N$(H$_2$) by the
cloud parameterization described in Sect.~\ref{sect_phys}), 
and $n_{\mathrm{fr}}$ is a free parameter 
that describes the characteristic freeze-out density of the species.
As shown in Table~\ref{tbl_abu}, all species in the sample can be fitted with
$n_{\mathrm{fr}}$ values in the range $(1-2) \times 10^5$~cm$^{-3}$,
characteristic of molecular freeze out.

For the freeze-out resistant N$_2$H$^+$ molecule, the above 
expression was substituted by 
$$f^\prime_{\mathrm{in}}(N({\mathrm{H}_2})) = 1 - e^{-(n(\mathrm{H}_2)/n_{\mathrm{fr}})^2},$$
which represents an abundance enhancement, and where $ n_{\mathrm{fr}}$ is 
again a free parameter that describes the characteristic density at which
the N$_2$H$^+$ enhancement occurs.

\begin{figure}
   \centering
   \includegraphics[width=\hsize]{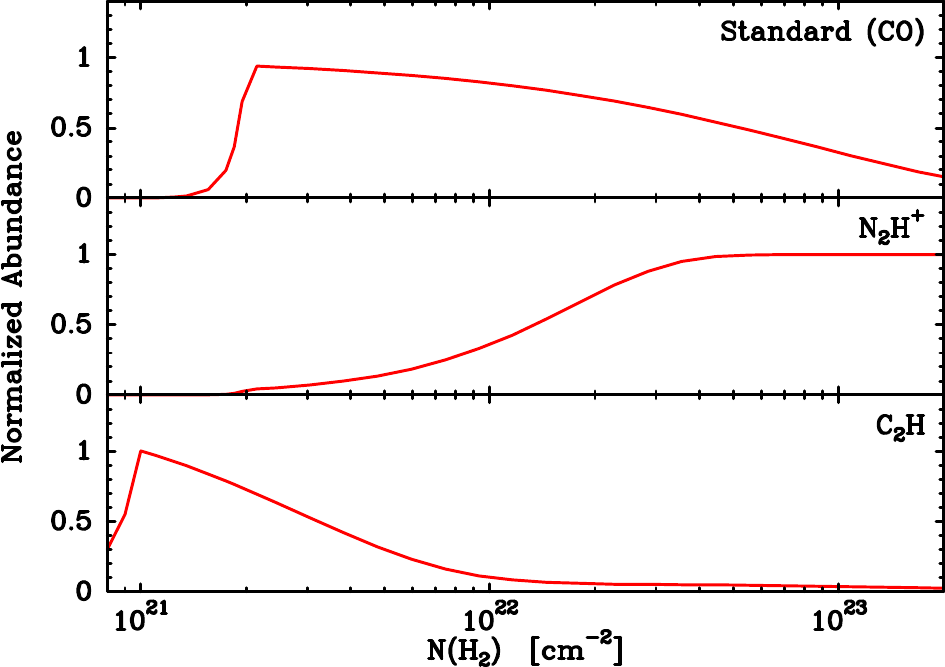}
   \caption{Three types of abundance profiles generated by a simple 
   	model and used to fit the emission from different species in Perseus.
   	{\em Top panel:} Standard abundance profile consisting of 
   	an outer photodissociation edge and an
   	inner freeze-out drop. It was used to fit most molecular data and is illustrated 
   	here with CO.
   	{\em Middle panel:} Profile with an inner enhancement required to fit the N$_2$H$^+$ data.
   	{\em Bottom panel:} Profile with an outer abundance enhancement 
   	required to fit the C$_2$H
   	and CN data (illustrated here with C$_2$H).}
   \label{abundances}
\end{figure}

Fig.~\ref{abundances} presents the three types of abundance profiles 
generated using our three-factor parameterization that were 
needed to fit the variety of intensity behaviors identified in our Perseus survey.
The top panel presents the abundance profile 
used to fit all species excluding N$_2$H$^+$, C$_2$H, and CN. 
This profile is
characterized by a sharp photodissociation edge at low column densities, 
a close-to-constant abundance value at intermediate column densities, 
and a gradual abundance drop due to freeze out at high column densities.
We refer to this profile as the ``standard'' one.
The middle panel shows the abundance profile needed to fit the 
N$_2$H$^+$ emission. It presents an opposite behavior to the standard 
profile at high column densities: an initial rapid abundance increase
followed by a close-to-constant value.
Finally, the bottom panel presents the abundance profile
used to fit the C$_2$H emission (a similar one was used for CN).
This profile presents a significant enhancement at low column densities
followed by an external photodissociation edge.

\section{Fitting results for remaining species}
\label{sect_remaining}

In Sect.~\ref{sect_mod_resu} we presented radiative transfer results
for a representative group of molecular species that cover the different 
emission behaviors found in Perseus:
the family of CO isotopologs, CS and HCN (as examples
of traditional dense gas tracers with standard abundance profile),
and N$_2$H$^+$ and C$_2$H, two species with inner and outer abundance enhanced
profiles, respectively. In this Appendix
we present the model results for the remaining
species of the survey.

As mentioned in Sect.~\ref{sect_mod_resu}, our model aims to fit 
simultaneously for each species
all the transitions of the main and
rare isotopologs detected in the 1mm, 2mm, and 3mm wavelength windows
covered by the survey. We assume that the abundances of the different
isotopologs are related by the ratios estimated
by \cite{wil94} in the solar vicinity.

\begin{figure*}
   \centering
   \includegraphics[width=\hsize]{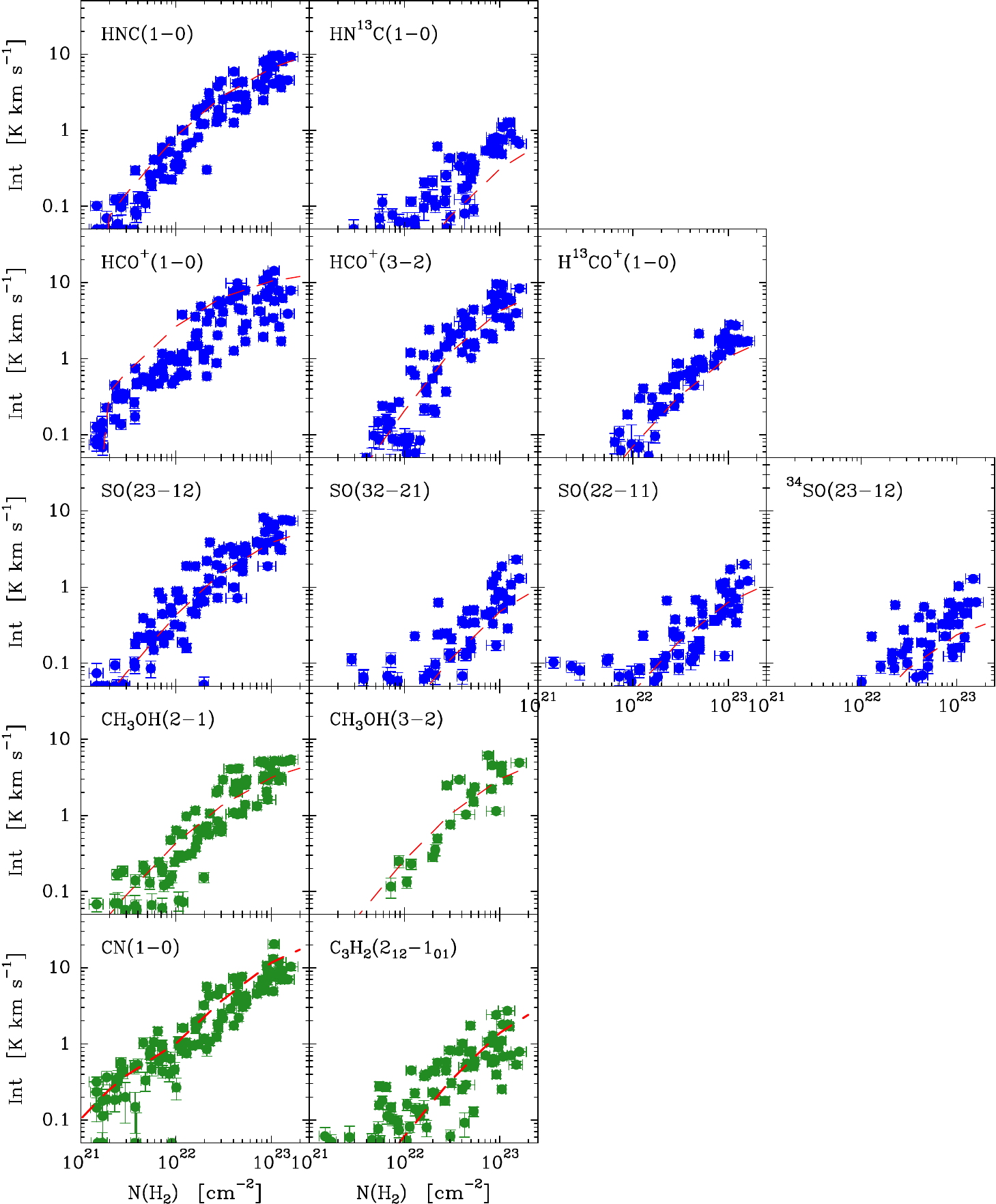}
   \caption{Comparison between observations and model results 
   for the species not presented in the main text. 
   The data points represent the observations (traditional dense gas tracers in blue and additional
   tracers in green), and the red dashed lines represent the model results.
   Each panel row contains all observed transitions and isotopologs of any given
   molecular species, except for the last row that shows together the only observed lines of 
   CN and C$_3$H$_2$.
   Several SO and C$_3$H$_2$ points can
   	be seen above the
   	plot lower limit at very low column densities. A visual 
   	inspection of their spectra suggests that they represent 
   	noise or baseline residuals, and not true molecular emission.
}
   \label{lvg_appendix}
\end{figure*}

Fig.~\ref{lvg_appendix} presents the fit results
for the tracers not presented in Sect.~\ref{sect_mod_resu}: the
traditional dense gas tracers HNC, HCO$^+$, and SO (blue symbols), and the
additional tracers CH$_3$OH, CN, and C$_3$H$_2$ (green symbols).
For each species, all observed transitions and isotopologs are
presented in the same row,
except for CN and C$_3$H$_2$, for which
only one transition was detected, and which are presented
together in the bottom row for space reasons.

As can be seen in the figure, the model predictions (dashed lines)
generally lie within the cloud of observed data points,
indicating that the model provides a reasonable fit to the observations.
Some deviations from the expected best fit are seen
in species where the lines span a large range of
optical depths, such as HNC and HCO$^+$.
These deviations are similar to those found 
when fitting the CS and HCN data, in the sense that
our model has difficulty reproducing simultaneously
the $J$=1--0 lines of the main and rare isotopologs
because our solution
slightly overestimates the thick main line
while it underestimates the thin isotopolog line.
As discussed in Sect.~\ref{sect_mod_resu}, this fitting
problem likely results from the use of
a single volume density
to characterize the physical conditions along any given line of sight,
a choice that
ignores the complex density structure present in the
real cloud.
Apart from this limitation, the model seems to work reasonably well
considering its small number of free parameters, the large number of
molecular species considered, and
the large dynamic range span by the intensities.
For these reasons we believe that the model captures the main features
of the physical and chemical structure of the Perseus molecular 
cloud.

Table~\ref{tbl_abu} summarizes the abundance best-fit parameters
used to model the emission from each survey species. It also provides 
references for the collision rate coefficients used to model the excitation.
As mentioned in Sect.~\ref{sect_mod_resu}, these coefficients, together
with any additional molecular parameters, were input into the radiative
transfer code using the files provided by the LAMDA database
\citep{sch05,van20}.

A significant result from Table~\ref{tbl_abu} is that
the abundance values derived for Perseus match closely abundance estimates
made for other star-forming regions. 
As mentioned before, the $A_0$ values of the photodissociation edge and the
characteristic freeze-out density $n_{\mathrm{fr}}$ agree with previous observations
and chemical model predictions for typical clouds 
(e.g., \citealt{ber97, cas99, ber02, taf02, rol07, pin08}).
In addition, the $X_0$ values in Table~\ref{tbl_abu} for all species 
with the exception of C$_3$H$_2$ agree within a factor 
of three with the geometrical mean of the undepleted abundances in the 
Taurus-Auriga L1498 and L1517B dense cores estimated by \cite{taf06} 
(the difference for C$_3$H$_2$ is a factor of 6). 
Considering that these two estimates used very different radiative transfer
assumptions and
even different collision rates, the agreement is notable. It suggests once again that
the gas properties derived from our Perseus analysis are likely representative of 
the gas properties in other star-forming regions.

\begin{table}
\caption[]{Best-fit abundance parameters and collision-rate coefficients \label{tbl_abu}}
\centering
\begin{tabular}{lccccc}
\hline
\noalign{\smallskip}
Species & $X_0$ & $A_0$ & $\alpha$ & $n_{\mathrm{fr}}$ & Coll. Coeff. \\
 & & (mag) & & (cm$^{-3}$) & Reference \\
\noalign{\smallskip}
\hline
\noalign{\smallskip}
CO & $9.5\; 10^{-5}$ & 2 & 0 & $10^5$ & 1 \\
HCN & $3\; 10^{-9}$ & 2 & 0 & $2\; 10^5$ & 2 \\
CS & $3\; 10^{-9}$ & 2 & 0 & $10^5$ & 3 \\
HNC & $10^{-9}$ & 2 & 0 & $2\; 10^5$ & 4 \\
HCO$^+$ & $1.5\; 10^{-9}$ & 2 & 0 & $2\; 10^5$ & 5 \\
SO & $10^{-9}$ & 2 & 0 & $2\; 10^5$ & 6 \\
CH$_3$OH & $1.5\; 10^{-9}$ & 2 & 0 & $2\; 10^5$ & 7 \\
CN & $1.5\; 10^{-9}$ & 1 & 5 & $2\; 10^5$ & 8 \\
C$_3$H$_2$ & $2\; 10^{-10}$ & 2 & 0 & $2\; 10^5$ & 9 \\
N$_2$H$^+$ & $1.5\; 10^{-10}$ & 2 & 0 & $3\; 10^4$ & 10 \\
C$_2$H & $2\; 10^{-9}$ & 1 & 20 & $2\; 10^5$ & 11 \\
\hline
\end{tabular}
\tablefoot{See Appendix~\ref{app_abu} for a full description of the abundance parameters.} 
\tablebib{
	(1)~\citet{yan10}; (2) \citet{her17}; (3) \citet{liq06}; (4) \citet{dum10};
	(5) \citet{flo99}; (6) \citet{liq07}; (7) \citet{rab10}; (8) \citet{kal15};
	(9) \citet{cha00}; (10) \citet{dan05}; (11) \citet{spi12}.
}
\end{table}

\end{document}